\documentclass[twocolumn]{aastex63}

\usepackage{amsmath}

\received{-}
\revised{-}
\accepted{-}

\submitjournal{APJ}

\shorttitle{Extreme r-process enhanced stars at high metallicity in Fornax}
\shortauthors{Reichert et al.}

\graphicspath{{./}{figures/}}

\begin{document}

\title{Extreme r-process enhanced stars at high metallicity in Fornax\footnote{Based on data obtained from the ESO reduced archive. Program IDs: 080.B-0784(A), 171.B-0588(A), 71.B-0641(A).}}

\correspondingauthor{M. Reichert}
\email{mreichert@theorie.ikp.physik.tu-darmstadt.de}

\author{M. Reichert}
\affiliation{Technische Universit\"at Darmstadt, Institut f\"ur Kernphysik, Schlossgartenstr. 2, 64289 Darmstadt, Germany \label{TUD}}

\author{C. J. Hansen}
\affiliation{ Max-Planck-Institut f\"ur Astronomie, K\"onigstuhl 17, D-69117 Heidelberg, Germany \label{MPI}}
\affiliation{ Copenhagen University, Dark Cosmology Centre, The Niels Bohr Institute, Vibenshuset, Lyngbyvej 2, DK-2100 Copenhagen, Denmark \label{Darkcosm}}

\author{A. Arcones}
\affiliation{Technische Universit\"at Darmstadt, Institut f\"ur Kernphysik, Schlossgartenstr. 2, 64289 Darmstadt, Germany \label{TUD}}
\affiliation{GSI Helmholtzzentrum für Schwerionenforschung GmbH, Planckstr. 1, D-64291 Darmstadt, Germany\label{GSI}}

\begin{abstract}
We present and discuss three extremely r-process enhanced stars located in the massive dwarf spheriodal galaxy Fornax. These stars are very unique with an extreme Eu enrichment ($1.25\le \mathrm{[Eu/Fe]} \le 1.45$) at high metallicities ($-1.3 \le \mathrm{[Fe/H]} \le -0.8$). They have the largest Eu abundances ever observed in a dwarf galaxy opening new opportunities to further understand the origin of heavy elements formed by the r-process. We derive stellar abundances of Co, Zr, La, Ce, Pr, Nd, Er, and Lu using 1-dimensional, local thermodynamic equilibrium (LTE) codes and model atmospheres in conjunction with state-of-the art yield predictions. We derive Zr in the largest sample of stars (105) known to date in a dwarf galaxy. Accurate stellar abundances combined with a careful assessment of the yield predictions have revealed three metal-rich stars in Fornax showing a pure r-process pattern. We define a new class of stars, namely Eu-stars, as r-II stars (i.e., [Eu/Fe]$>1$) at high metallicities (i.e., $\mathrm{[Fe/H]}\gtrsim -1.5$). The stellar abundance pattern contains Lu, observed for the first time in a dwarf galaxy, and reveals that a late burst of star formation has facilitated extreme r-process enhancement late in the galaxy's history ($<4$\,Gyr ago). Due to the large uncertainties associated with the nuclear physics input in the yield predictions, we cannot yet determine the r-process site leading to the three Eu-stars in Fornax. Our results demonstrate that extremely r-rich stars are not only associated with ultra faint low-mass dwarf galaxies, but can be born also in massive dwarf galaxies.

\end{abstract}


\section{Introduction} \label{sec:intro}
The creation of heavy elements poses a number of open and interesting questions going from  a small scales to the largest scales including formation of stars or even galaxies. Recent studies of Milky Way (MW) halo stars  have  shown that stars rich in neutron-capture elements may have originated in now dissolved (ultra faint) dwarf galaxies \citep{Roederer2018b}. Limited to smaller sample sizes, the results seem to indicate that stars with the strongest rapid neutron-capture process (r-process) enhancement have formed ex situ and later been accreted into the MW. This makes dwarf galaxies excellent study cases for understanding the r-process.

Fornax (Fnx) is one of the most massive dwarf spheriodals (dSph) in the Local Group (LG). It shows a unique star formation history with a sudden increase of star formation only $\sim4$\,Gyr ago \citep{deBoer2012, Lemasle2014,Hendricks2014,Weisz2014,Rusakov2021}. Furthermore, Fornax might currently have gas left, however, this is still under debate \citep{Bouchard2006}. All dwarf galaxies show individual trades and chemical imprints \citep{Grebel1997,Tolstoy2009}.

Here we present three Fornax stars with extreme high Eu abundances (rII stars, [Eu/Fe]\footnote{$\mathrm{[Eu/Fe]} = \log \left( N_\mathrm{Eu} / N_\mathrm{Fe} \right)-\log \left( N_\mathrm{Eu} / N_\mathrm{Fe} \right)_\odot$, with the number of europium and iron atoms per cm$^3$, $N_\mathrm{Eu}$ and $N_\mathrm{Fe}$.}$>1$, \citealt{Beers2005}) at high metallicities. Moreover, we report a large study of the neutron-capture element Zr in the dSph Fornax. Our study  presents the first detection of the heavy element Lu in a dSph. In contrast to most r-process enriched stars, typically found at low-metallicities ([Fe/H]$\sim -2$), the three Eu-stars in Fornax are relatively metal-rich ([Fe/H]$\sim -1$). Hence, we use them to understand the astrophysical source of the r-process by comparing to various yield predictions. 

Stars with  peculiar enhanced r-process abundances have rarely been observed at these high metallicities. Most of them are clearly members of a dSph galaxy, e.g., Ursa Minor (Cos 82, \citealt{Shetrone2001,Aoki2007,Sadakane2004} or Sculptor, SCMS 982, \citealt{Geisler2005,Skuladottir2020b}). We note that there are also neutron-capture enhanced stars at similar metallicities reported already by \citet{Letarte2010} in Fornax, which were not found in \citet{Letarte2018} and \citet{Reichert2020}. Also MW field stars, for example, 2MASS 18174532-3353235 \citep{Johnson2013} a possible bulge contender, shows a high neutron-capture enhancement at a high metallicity. Moreover, some halo stars show similar abundance enhancements and are thought to be accreted into the MW halo, as indicated by a low $[\alpha/\mathrm{Fe}]$ ratio (J1124+4535, \citealt{Xing2019}), while for others, their origin is, however, not so clear. Also HD 222925 is r-rich and located in the halo, but due to its highly eccentric and retrograde orbit it has been suggested to be accreted from a satellite galaxy \citep{Roederer2018b}. By looking at the alpha-abundances and fitting a knee to the stellar population, the dSph mass can be assessed. Following the fit of the $\alpha$-knee of \citet[][Eq.~6]{Reichert2020}, the host environment of this star must have at least the size of Sagittarius or Fornax and an accretion scenario may therefore be possible \citep[see also][for a discussion on the origin of HD 222925]{Roederer2018}. 

The astrophysical sites of the r-process are still under discussion, even if recent kilonova \citep{Abbott2017,Watson2019} has shown that  neutron star mergers (NSM) can produce heavy elements. Galactic chemical evolution (GCE) models \citep[e.g.,][]{Matteucci2014, Cote2019, Kobayashi2020} suggest that an additional site may be active in our galaxy. An exciting possibility are  magneto-rotational supernovae \citep[MR-SNe;][]{Winteler2012,Nishimura2015, Nishimura2017, Moesta2018, Reichert2021}. Stars at intermediate metallicities are highly mixed and remain poorly studied in the GCE scheme. Also collapsars \citep{Siegel2019, Miller2019, Just2021} have been discussed as a potential site. Our results indicate that both sites, NSM and MR-SNe, can explain the Eu-stars and probe a late star formation burst in Fornax.

This paper is organised as follows. In Sect.~\ref{sec:obs}, we present the observations, sample, and stellar parameters. The abundances are shown in Sect.~\ref{sec:ab} including the largest Zr sample in a dwarf galaxy and heavy neutron-capture elements with focus on Eu and the first Lu detection in a dwarf galaxy. In Sect.~\ref{sec:origin}, we discuss in detail the possible origin of the Eu-stars, which is associated with one r-process event and requires late star formation. Finally, we conclude in Sect.~\ref{sec:conclusions}.

\section{Observations, sample, and stellar parameters}
\label{sec:obs}

\subsection{Observations}
We present four Fornax stars including three with anomalous high Eu ($1.25\le \mathrm{[Eu/Fe]} \le 1.45$). 
The quality of the available observations allows for an accurate and precise determination of abundances (typically to within $\pm0.2$\,dex). The stars were observed at high and low resolution with FLAMES/GIRAFFE \citep{FLAMES} with the setups HR10, HR13, HR14, and LR8  (see Table~\ref{tab:grating}). The LR8 setup \citep{Battaglia2006} is used here only for radial velocity determinations.  We use reduced spectra from the ESO archive\footnote{https://archive.eso.org/wdb/wdb/adp/phase3\_main/form} and perform the data processing (sky correction, radial velocity shifts, co-adding, and normalisation of the spectra) as in \citet{Reichert2020}.

\begin{center}  
\begin{table}
\caption{Different FLAMES/GIRAFFE setups showing minimum and maximum wavelength coverage and resolving power ($R$).}            
\label{tab:grating}     
\begin{tabular}{l c c c }      
\hline\hline                
Grating & $\lambda_{min} [\mathrm{\AA}]$& $\lambda_{max} [\mathrm{\AA}]$ & $R$ \\    
\hline                      
HR10  &  5339 & 5619 & 19800 \\
HR13  &  6120 & 6405 & 22500 \\
HR14  &  6308 & 6701 & 17740 \\
LR8   &  8206 & 9400 & 6500  \\
\hline                                 
\end{tabular}
\end{table}
\end{center} 

\subsection{Sample}
\label{sec:sample}
   \begin{figure}[h]
   \centering
   \includegraphics[width=\hsize]{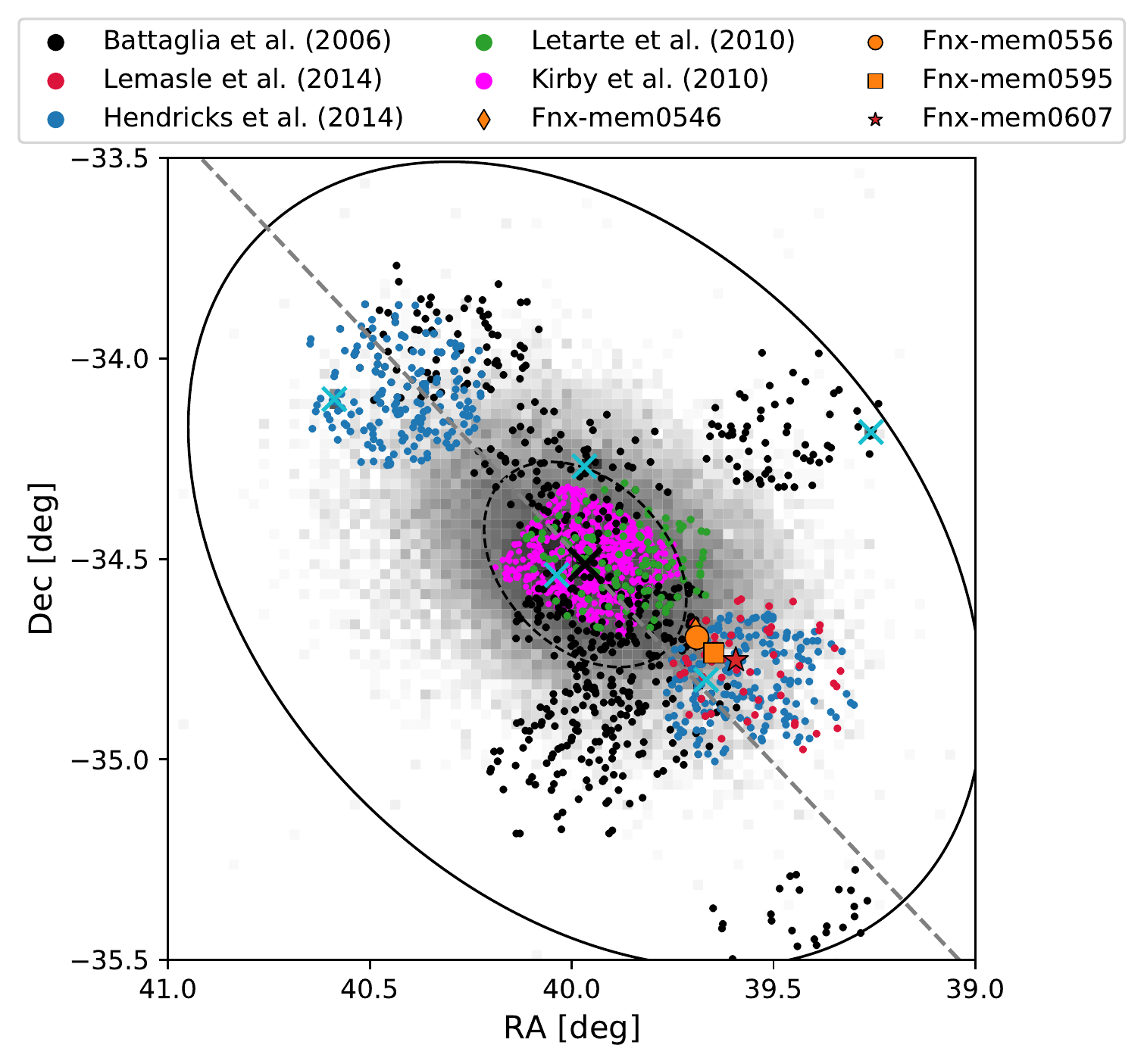}
      \caption{Coordinates of sample stars together with pointings of \citet{Battaglia2006}, \citet{Lemasle2014}, \citet{Hendricks2014}, \citet{Letarte2010}, and \citet{Kirby2010}. The large oval shows the nominal tidal radius and the small dashed oval the core radius \citep{Battaglia2006}. The black cross marks the center of the galaxy as given by \citet{Battaglia2006}. The grey 2D histogram shows the amount of stars observed by \textit{Gaia} \citep[DR2,][]{Gaia2018,Gaia_vizier2018}. Cyan crosses indicate the globular clusters of Fornax \citep{Mackey2003}.} 
         \label{fig:location}
   \end{figure}

Our sample consists of Fornax stars from \citealt{Reichert2020} who analysed $380$ stars in dSph galaxies. The stars have been extracted from the reduced ESO and KOA archives. More specifically, we assigned a star membership if its coordinates agree within a circle of three arcsec with the coordinates in the SIMBAD \citep{Simbad}, NED \citep{Mazzarella2007}, or the second {\it Gaia} data release \citep[DR2,][]{Gaia2018,Gaia_vizier2018}.

Therefore, the three Eu-enhanced stars are clearly members of Fornax as previous studies have already assigned them Fornax membership based on radial velocities, and distance to the centre of Fornax \citep[see Fig.~\ref{fig:location} and ][]{Battaglia2006,Lemasle2014}. Also the proper motions of \textit{Gaia} \citep[DR2,][]{Gaia2018,Gaia_vizier2018} indicate their membership (see Table~\ref{tab:prop_motions}).

  \begin{table}[h]
\caption{Proper motions in right ascension direction $\mu _\delta$, and in declination direction $\mu _\alpha \cos \delta$ taken from \textit{Gaia} \citep[DR2,][]{Gaia2018,Gaia_vizier2018}. Furthermore, we list the mean values of Fornax \citep{McConnachie2020}.}        
\label{tab:prop_motions}     
\centering                   
\begin{tabular}{l c c }      
\hline\hline               
Object & $\mu _\delta$ [$\mathrm{mas \, yr^{-1}}$]& $\mu _\alpha \cos \delta$ [$\mathrm{mas \, yr^{-1}}$] \\   
\hline                       
Fnx-mem0546   &  $0.088 \pm 0.145$ & $ -0.199 \pm 0.210$  \\
Fnx-mem0556   &  $0.400 \pm 0.148$ & $-0.775 \pm 0.203$ \\
Fnx-mem0595   &  $0.344 \pm 0.120$ &  $-0.237 \pm 0.167$ \\
Fornax (mean) &  $0.038 \pm 0.003$ & $-0.416 \pm 0.004$  \\
\hline                                  
\end{tabular}
\end{table}
   
\begin{figure}[h]
   \centering
   \includegraphics[width=\hsize]{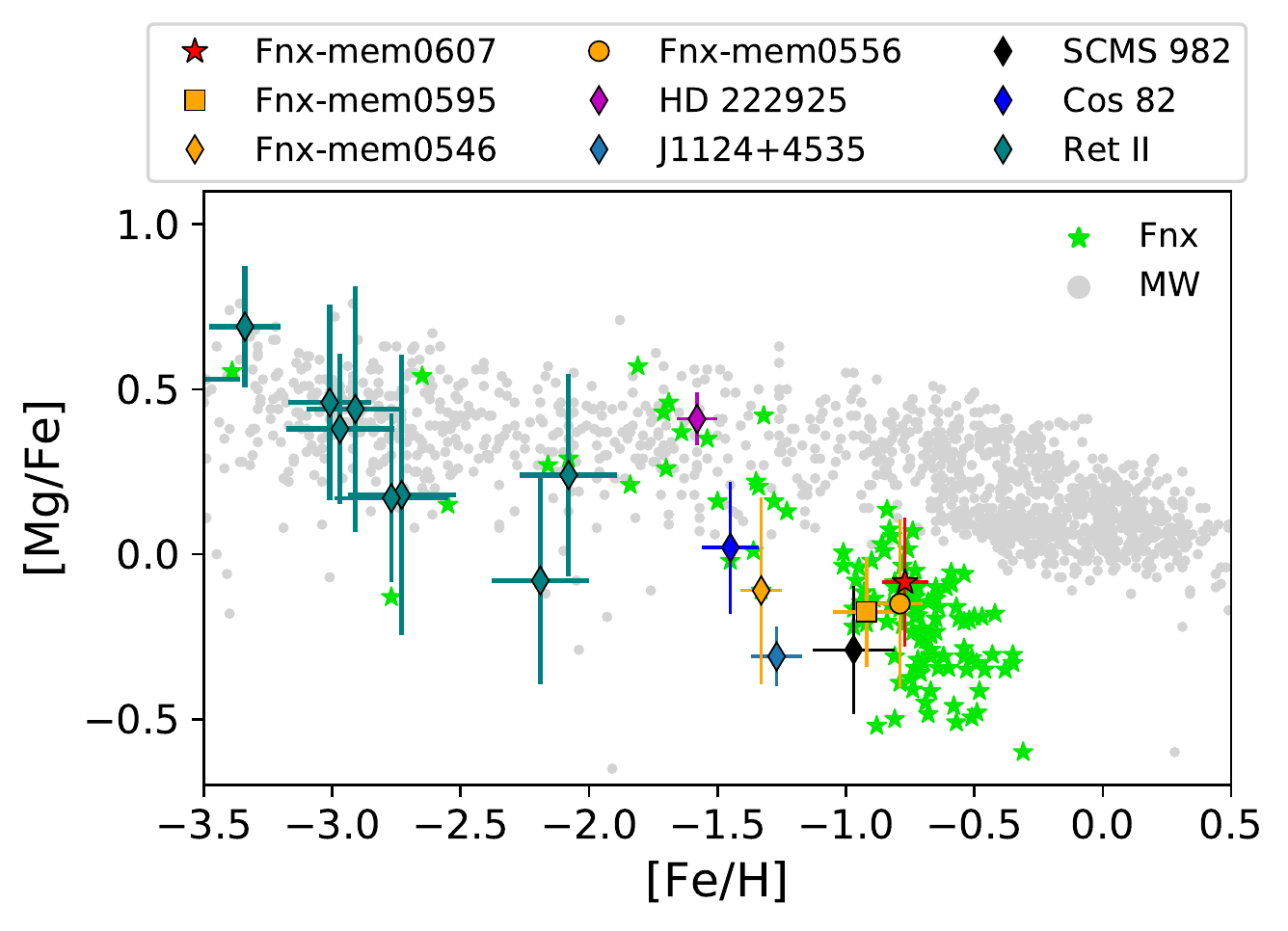}
      \caption{[Mg/Fe] versus metallicity is shown for the four Fornax stars that we investigate here \cite[see also][]{Reichert2020}:  Fnx-mem0607 (red star), Fnx-mem0595 (yellow square), Fnx-mem0546 (yellow diamond), and Fnx-mem0556 (yellow circle). For comparison we show other Fornax stars   \citep[green stars, ][]{Reichert2020} and MW stars \cite[grey dots, ][]{Reddy2003, Cayrel2004, Reddy2006, Ishigaki2013, Fulbright2000, Nissen1997, Prochaska2000,  Stephens2002, Ivans2003, McWilliam1995,  Ryan1996, Gratton1988, Edvardsson1993, Roederer2016, Ezzeddine2020}. Moreover, we compare to other typical r-process enhanced stars:  Cos 82 \citep[blue diamond,][]{Shetrone2001}, SCMS 982 \citep[black diamond,][]{Geisler2005}, HD 222925 \citep[magenta diamond,][]{Roederer2018}, J1124+4535 \citep[blue diamond,][]{Xing2019}, and Reticulum II \citep[teal diamonds,][]{Ji2016b}. In all following figures we keep the same symbols and colors for the various stars.
} 
         \label{fig:mg_evolution}
   \end{figure}

The [$\alpha$/Fe] ratio of the stars also fits with the general trend of Fornax (Fig.~\ref{fig:mg_evolution}) and similar to many other r-process enriched metal-rich stars, a clear contribution from type Ia SNe is visible. 

\subsection{Stellar parameters}
The stars were previously analysed by \citet{Lemasle2014}, \citet{Reichert2020}, and \citet{Battaglia2006}. Table~\ref{tab:stel_pars} shows their determined effective temperature, surface gravity, and metallicity (for \citealt{Battaglia2006} the metalicities are from the calcium triplet).
 We note that the star Fnx-mem0546 was excluded from  \citet{Lemasle2014} due to convergence problems when determining the stellar parameters. 
Here, we adopt stellar parameters from \citet{Reichert2020}. The effective temperatures and metallicities were derived using the automatic code SP\_Ace \citep{Boeche2016} which fits a theoretically calculated spectrum to the observed spectrum of the star until convergence is reached. The theoretical spectrum is constructed by using a previously calculated curve of growth library. We note that the metallicity is therefore also biased by the LTE assumptions.
 \newline
 The surface gravities were derived using: 
 \begin{align} \log g_* = \log g_{\odot} +\log \frac{M_*}{M_\odot} + 4\, \log \frac{T_{\text{eff},*}}{T_{\text{eff},\odot}}\\ +0.4\left( M_{\text{bol},*} - M_{\text{bol},\odot} \right), \nonumber \end{align} with $\log g_{\odot}=4.44$, $T_{\text{eff},\odot}=5780\, \rm K$, $M_*=0.8\pm 0.2 M_\odot$ and $M_{\text{bol},\odot}=4.72\, \rm mag$, and the distance to Fornax of $147\pm8 $\, kpc taken from \citealt{Karczmarek2017}).
 \newline
The microturbulence was calculated with an empirical relation from \citet{Kirby2010}:
\begin{equation}
    \xi _\mathrm{t} = ((2.13\pm 0.05) - (0.23\pm 0.03)\cdot \log g) \, \rm km \, s^{-1}.
\end{equation}
In general, effective temperatures in \citet{Reichert2020} are hotter than in \citet{Lemasle2014} because of the different methods in deriving them (spectroscopic versus photometric, see Appendix~B of \citealt{Reichert2020} for a more detailed discussion). However, we stress that the strong neutron-capture element enhancement  is also present when using the stellar parameters of \citet{Lemasle2014}, hence the choice of stellar parameters does not change our conclusions.
\begin{table*}
\caption{Stellar parameters from different studies}         
\label{tab:stel_pars}      
\centering                        
\begin{tabular}{c c c c c c c c}       
\hline\hline                
& \multicolumn{3}{c}{\citet{Reichert2020}} & \multicolumn{3}{c}{\citet{Lemasle2014}} & \citet{Battaglia2006}\\
Identifier & $T_\mathrm{eff}$ & $\log g$ & [Fe/H] & $T_\mathrm{eff}$ & $\log g$ & [Fe/H] & [Fe/H] \\  
\hline                   
Fnx-mem0546 & $4367\pm 110$ & $0.82 \pm 0.13$ & $-1.33 \pm 0.08$ & - & - & -&$-1.31\pm 0.07$\\
Fnx-mem0556 & $4176\pm 58$  & $0.70 \pm 0.12$ & $-0.79 \pm 0.09$ & $3971 \pm 150$ & $0.64 \pm 0.5$ & $-0.63 \pm 0.03$ & $-0.64 \pm 0.16$\\
Fnx-mem0595 & $4223\pm 67$  & $0.66 \pm 0.12$ & $-0.92 \pm 0.13$ & $3889 \pm 150$ & $0.51 \pm 0.5$ & $-0.88 \pm 0.03$ & $-1.04 \pm 0.08$\\
Fnx-mem0607 & $4207\pm 63$  & $0.69 \pm 0.12$ & $-0.77 \pm 0.09$ & $3916 \pm 150$ & $0.55 \pm 0.5$ & $-0.96 \pm 0.02$ & -\\
\hline                                  
\end{tabular}
\end{table*}

\section{Abundances}
\label{sec:ab}

We combine the previously determined abundances \citep{Reichert2020} with  eight newly analyzed elements: Co, Zr, La, Ce, Pr, Nd, Er, and Lu (Table~\ref{tab:linelist}). Hyperfinesplitting is included for the redder lines of La, Co, and Lu \citep{Lawler1989,Kurucz2011}. We derive all additional abundances by synthesizing and fitting theoretical spectra with the local thermodynamic equilibrium (LTE) spectrum synthesis code MOOG \citep[version of 2014,][]{Moog} using 1-dimensional model atmospheres from \citet{Kurucz1970}\footnote{http://kurucz.harvard.edu/grids.html}, and solar abundances from \citet{Asplund2009}. In addition, we have tried to extract carbon from several  molecular absorption lines but those  were too weak in the covered spectral region. Therefore, we can  exclude a strong carbon enhancement.

\begin{table}
\caption{Lines analyzed in addition to those used in \citet{Reichert2020}.}     \label{tab:linelist}    
\centering
\begin{tabular}{l l c c c }        
\hline\hline                
Element & Wavelength& EP& $\log gf$ & Literature\\    
\hline                        
Co I     &   $5483.34$ & 1.709  & -1.500 & 1\\
Y II     &   $5509.89$ & 0.992  & -1.015 & 2\\
Zr I     &   $6127.46$ & 0.154  & -1.060 & 3\\
Zr I     &   $6134.62$ & 0.000  & -1.280 & 3\\
Zr I     &   $6143.20$ & 0.071  & -1.100 & 3\\
La II    &   $6320.43$ & 0.173  & -1.610 & 4\\
La II    &   $6390.46$ & 0.321  & -1.410 & 5\\
Ce II    &   $5512.08$ & 1.007  & -0.390 & 6\\
Pr II    &   $5509.15$ & 0.482  & -1.168 & 7\\
Pr II    &   $6165.94$ & 0.923  & -0.205 & 7\\
Nd II    &   $5356.97$ & 1.263  & -0.280 & 8\\
Nd II    &   $5361.17$ & 0.559  & -1.480 & 8\\
Nd II    &   $5361.47$ & 0.680  & -0.370 & 7\\
Er II    &   $5414.60$ & 0.000  & -2.499 & 7\\
Lu II    &   $6221.87$ & 1.540  & -0.760 & 9\\
\hline                                   
\end{tabular}
\tablerefs{(1) \citet{Lawler2015}; (2) \citet{Hannaford1982}; (3) \citet{Biemont1981}; (4) \citet{Corliss1962}; (5) \citet{Lawler2001b}; (6) \citet{Lawler2009}; (7) \citet{Meggers1975}; (8) \citet{Denhartog2003}; (9) \citet{denhartog1998}.}
\end{table}

All derived abundances, including those already obtained by \citet{Reichert2020}, are presented in Table~\ref{tab:abundances} for the four Fornax stars. The star Fnx-mem0607 is used as a reference because it has typical abundances of neutron-capture elements in Fornax stars. The stars Fnx-mem0556, and Fnx-mem0595 have the highest europium abundance ever observed to our knowledge with $\log \epsilon (\mathrm{Eu})=0.98$ (Table~\ref{tab:abundances}). Within uncertainties, this is comparable only to SCMS~982 in Sculptor (\mbox{$\log \epsilon (\mathrm{Eu})=0.95 \pm 0.18$}, \citealt{Geisler2005}). In the following, we will refer to them as Eu-stars, defining them as r-II stars (i.e., [Eu/Fe]$>1$) at high metallicities (i.e., [Fe/H]$\gtrsim -1.5$).

\begin{table*}
\caption{Derived absolute abundances (including also those already presented in \citet[][R20]{Reichert2020}).}
\label{tab:abundances}      
\centering                          
\begin{tabular}{l c c c c r}        
\hline\hline                 
Element & Fnx-mem0546 & Fnx-mem0556 & Fnx-mem0595 & Fnx-mem0607 & Reference\\ 
\hline                        
Mg I & $6.16 \pm 0.27$ & $6.66 \pm 0.24$ & $6.51\pm 0.11$ & $6.75 \pm 0.17$ & R20 \\  
Sc I & $1.26 \pm 0.26$ & $1.79 \pm 0.10$ & $1.60\pm 0.29$ & $1.93 \pm 0.11$ & R20\\    
Ti II& $3.93 \pm 0.31$ & $4.32 \pm 0.25$ & $3.79\pm 0.28$ & $4.08 \pm 0.11$ & R20\\
Cr I & $3.57 \pm 0.39$ & $5.22 \pm 0.26$ & $4.76\pm 0.18$ & $4.86 \pm 0.44$ & R20\\
Mn I & $3.90 \pm 0.37$ & $3.97 \pm 0.14$ & $4.15\pm 0.23$ & $4.31 \pm 0.28$ & R20\\
Fe I & $6.17 \pm 0.08$ & $6.71 \pm 0.09$ & $6.58\pm 0.13$ & $6.73 \pm 0.09$ & R20\\  
Co I & $3.29 \pm 0.29$ & $3.90 \pm 0.26$ & $3.59\pm 0.24$ & $3.93 \pm 0.19$ & This study\\  
Ni I & $4.76 \pm 0.20$ & $5.14 \pm 0.21$ & $5.12\pm 0.23$ & $5.11 \pm 0.23$ & R20\\ 
Y II & $0.89 \pm 0.25$ & $1.35 \pm 0.29$ & $1.11\pm 0.24$ & - & R20\\ 
Zr  I & $2.19 \pm 0.30$ & $2.29 \pm 0.13$ & $2.12\pm 0.18$ & $1.70 \pm 0.20$ & This study \\ 
Ba II& $1.25 \pm 0.12$ & $1.76 \pm 0.30$ & $1.79\pm 0.14$ & $1.65 \pm 0.12$ & R20\\
La II& $1.50 \pm 0.28$ & $1.47 \pm 0.26$ & $1.04\pm 0.13$ & $0.60 \pm 0.09$ & This study\\
Ce II& $1.03 \pm 0.25$ & $1.33 \pm 0.24$ & $1.45\pm 0.24$ & - &  This study\\
Pr II& $1.17 \pm 0.25$ & $1.15 \pm 0.34$ & $1.01\pm 0.23$ & $0.33 \pm 0.15$ & This study\\
Nd II& $1.49 \pm 0.19$ & $1.89 \pm 0.29$ & $1.82\pm 0.30$ & $0.99 \pm 0.22$ & This study\\
Eu II& $0.64 \pm 0.19$ & $0.98 \pm 0.13$ & $0.98\pm 0.12$ & $0.18 \pm 0.34$ & R20\\
Er II& -               & $1.69 \pm 0.25$ & $1.53\pm 0.28$ & - & This study\\
Lu II& $0.22 \pm 0.23$ & $0.61\pm 0.26$         & $0.18\pm 0.37$        & $<0.23$ & This study \\
 \hline    
 $[$Ba$/$Eu$]$ & $-1.05$ $(r)$ &  $-0.88$ $(r)$ &  $-0.85$ $(r)$ & $-0.19$ $(r+s)$& -\\
  $[$Ba$/$La$]$ & $-1.33$ $(r)$ &  $-0.80$ $(r)$ &  $-0.33$ $(r)$ & $-0.03$ $(r+s)$ & -\\
 \hline\hline
\end{tabular}
\end{table*}   

\begin{figure}[h]
  \centering
  \includegraphics[width=\hsize]{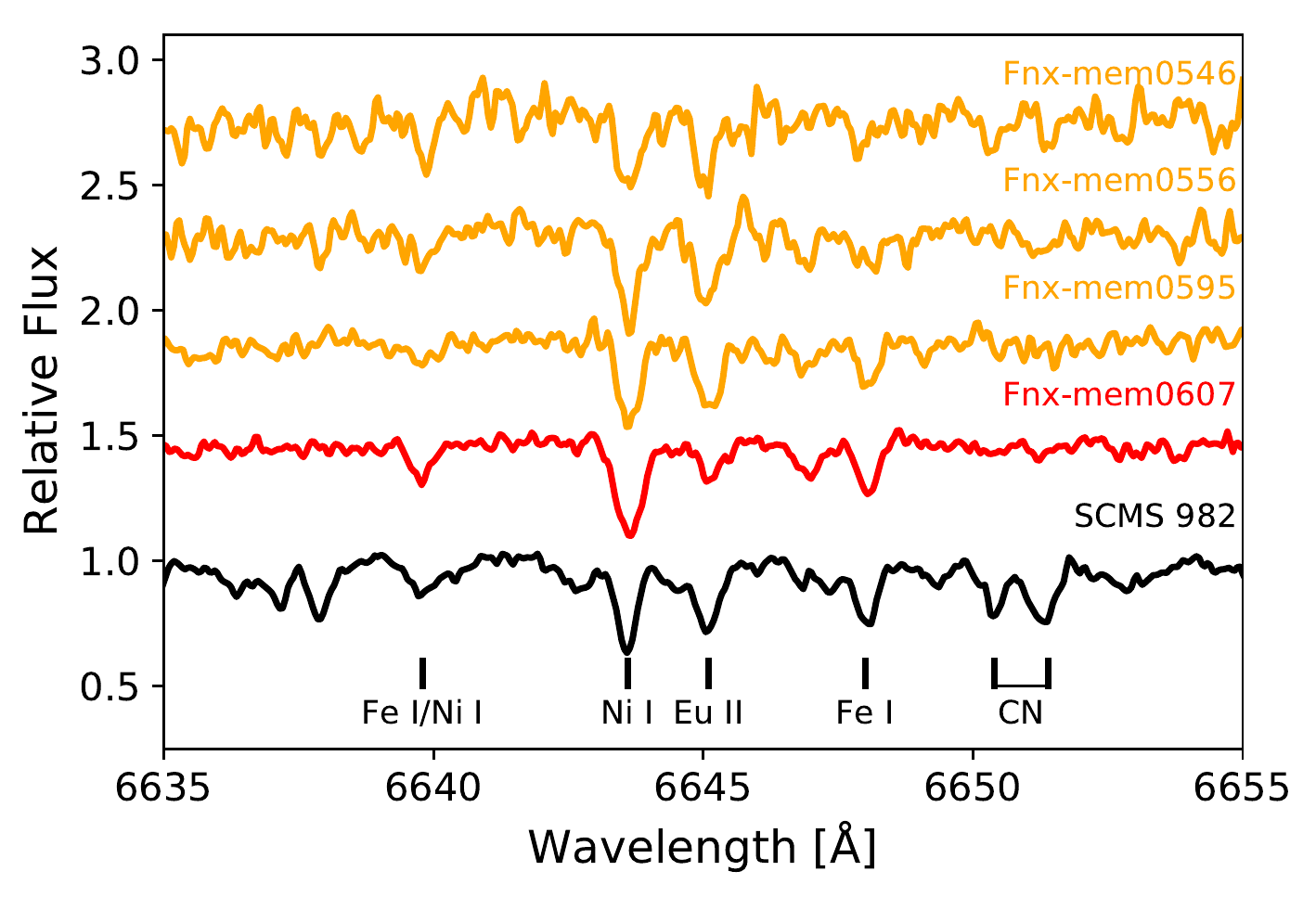}\\
  \includegraphics[width=\hsize]{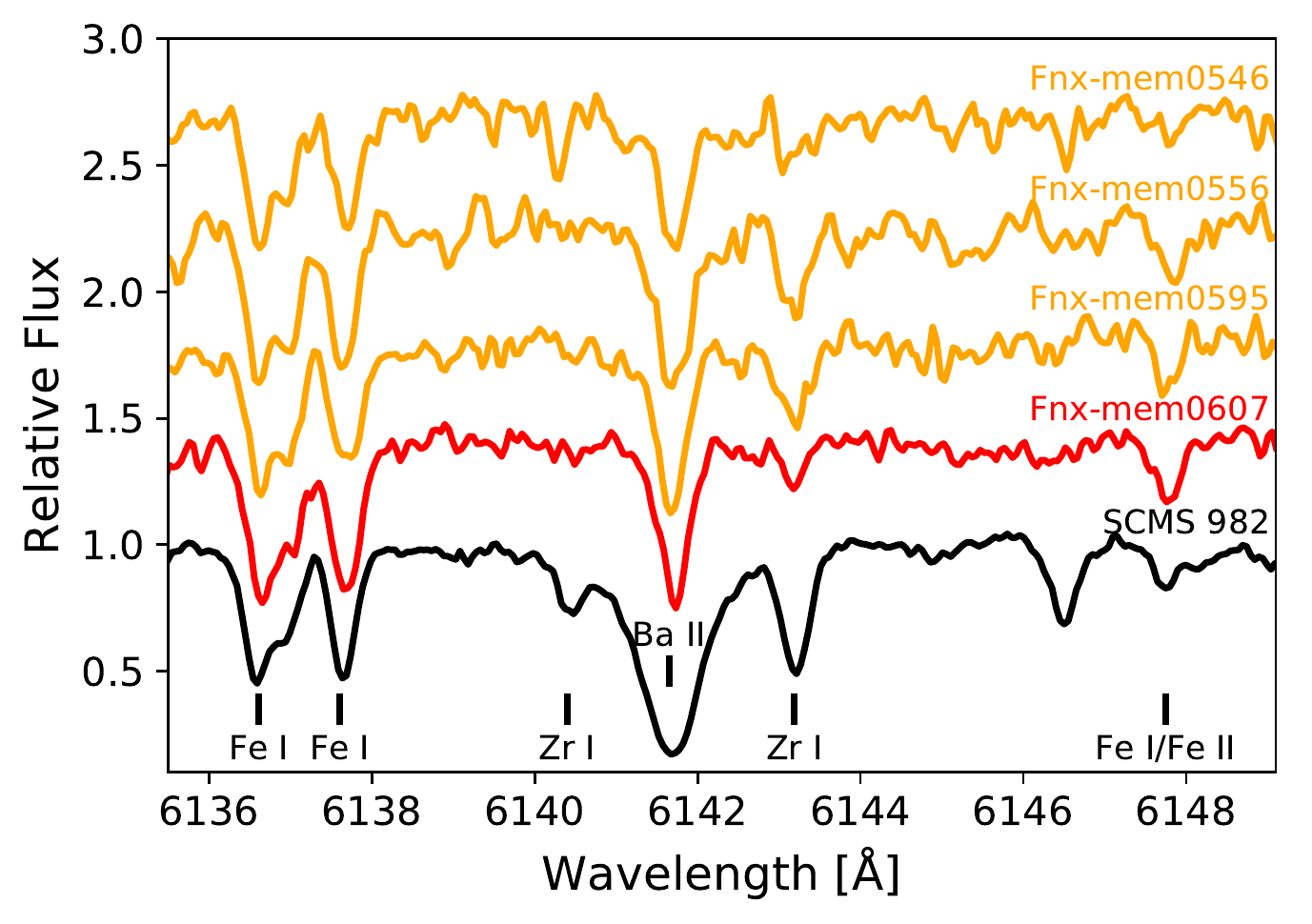}
\caption{Spectra (including an offset for better visibility) for Fnx-mem0546, Fnx-mem0556, Fnx-mem0595, Fnx-mem0607, and SCMS~982 for regions around the Eu absorption line (top panel) and around a Ba II line (bottom panel).}
\label{fig:spectra}
  \end{figure}

The enhancement of neutron capture elements (Zr, Ba, and Eu) is evident from the spectra shown in Fig.~\ref{fig:spectra}, where we include also SCMS~982 \citep{Geisler2005}, a star with similar absolute europium abundances. All  stars have similar stellar parameters and the strength of the lines can therefore be roughly compared (Fnx-mem0546 is slightly more metal-poor than the other stars). The europium absorption line is strongest for Fnx-mem0546, Fnx-mem0556, and Fnx-mem0595  and even comparable in strength to the nickel line close by. Note that SCMS~982 has  stronger Ba and Zr lines than the Fornax stars indicating a considerable s-process or i-process enhancement of this Sculptor star (see Sect.~\ref{sec:pattern}).

In the following, we will compare the four Fornax stars to other stars keeping symbols and colors the same in all figures. Fnx-mem0546, Fnx-mem0556, Fnx-mem0595, Fnx-mem0607 \citep{Reichert2020} are marked  by an orange diamond, a circle, a square, and a red star, respectively. Other Fornax stars  \citep{Reichert2020} are shown as green stars, and the MW stars as grey dots (\citet{Reddy2003}, \citet{Cayrel2004}, \citet{Reddy2006}, \citet{Ishigaki2013}, \citet{Fulbright2000}, \citet{Nissen1997}, \citet{Prochaska2000}, \citet{Stephens2002}, \citet{Ivans2003}, \citet{McWilliam1995}, \citet{Ryan1996}, \citet{Gratton1988}, \citet{Edvardsson1993}, \citet{Roederer2016}, \citet{Hansen2012} and \citet{Ezzeddine2020}). Moreover, we show other typical r-process enhanced stars:  Cos~82 \citep[blue diamond,][]{Shetrone2001}, SCMS~982 \citep[black diamond,][]{Geisler2005}, HD 222925 \citep[magenta diamond,][]{Roederer2018}, J1124+4535 \citep[blue diamond,][]{Xing2019}, and Reticulum II \citep[teal diamonds,][]{Ji2016b}.

\subsection{Largest Zr sample in a dwarf galaxy}
\label{sec:Zr}

\begin{figure}[h]
    \centering
    \includegraphics[width=\hsize]{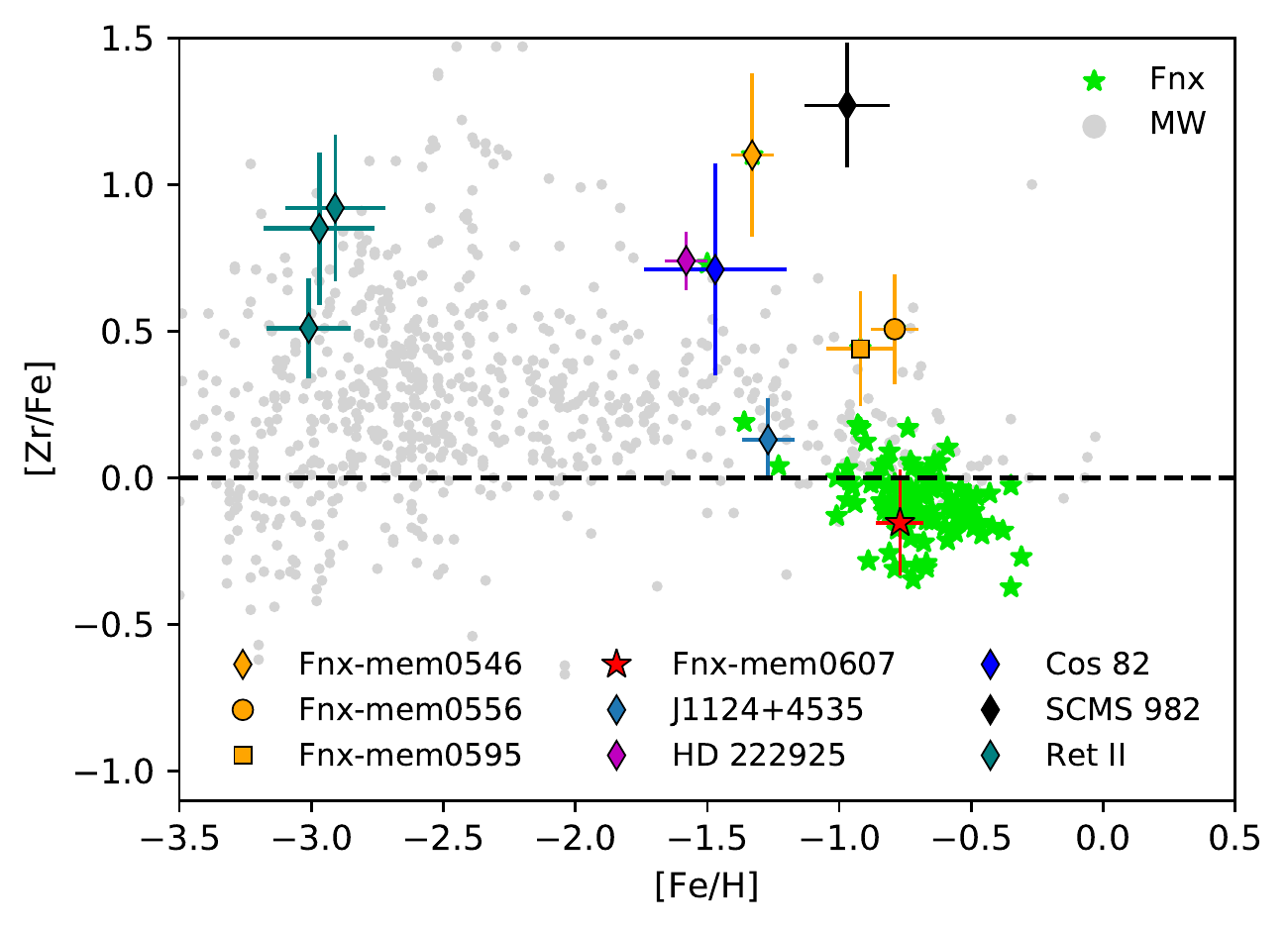}
    \caption{[Zr/Fe] versus metallicity following the same notation and references as in Fig.~\ref{fig:mg_evolution}.}
    \label{fig:zr_evolution}
\end{figure}

We have derived (neutral) zirconium abundances for all Fornax stars presented in \citet{Reichert2020}.  For many stars it is the first Zr determination, making it the largest sample of Zr abundances in a dwarf galaxy to date. Figure~\ref{fig:zr_evolution} shows the $\mathrm{[Zr/Fe]}$ in Fornax  including the neutron-capture enhanced stars. The average Zr abundance for all Fornax stars is slightly sub solar and the reference star (Fnx-mem0607) is located slightly below this general trend. There may be corrections to the abundances due to the LTE assumption which tends to affect neutral atoms more when they are the minority species as it is the case for Zr \citep[c.f., ][for a discussion]{Andrievsky2017}. 
The exact size of the NLTE correction depends on several factors such as the stellar parameters and the abundance value. Currently, there is no large, available Zr NLTE grid covering our parameter space. However, based on NLTE computations in \citet{Velichko2010} and more recent ones in \citet{Hansen2020} we would estimate corrections of the order of $\sim0.0$ to $>+0.3$\,dex depending on the Zr abundances and stellar parameters. For our three enhanced stars the corrections would likely amount to $\gtrsim+0.3$\,dex.

The three Eu-enhanced Fornax stars (Fnx-mem0546, Fnx-mem0556, and Fnx-mem0595) are clearly enhanced also in Zr. This agrees with these stars having an r-process pattern typical of other r-I or r-II stars, see for example the Reticulum II stars (teal diamonds) in Fig.~\ref{fig:zr_evolution} and the discussion in Sect.~\ref{sec:pattern}. Despite the high metallicities and Zr being dominated by the s-process in the solar system ($\sim 66\%$, \citealt{Bisterzo2014}); at low metallicities Zr is formed by the r-process as shown by most r-II stars. For the Eu-enhanced stars in Fornax, the enhancement of Zr is probably produced by an r-process event and not by the s-process, see Sect.~\ref{sec:pattern}.

\subsection{Heavy neutron-capture elements}
\label{sec:heavy_ncap}

\begin{figure}
   \centering
   \includegraphics[width=\hsize]{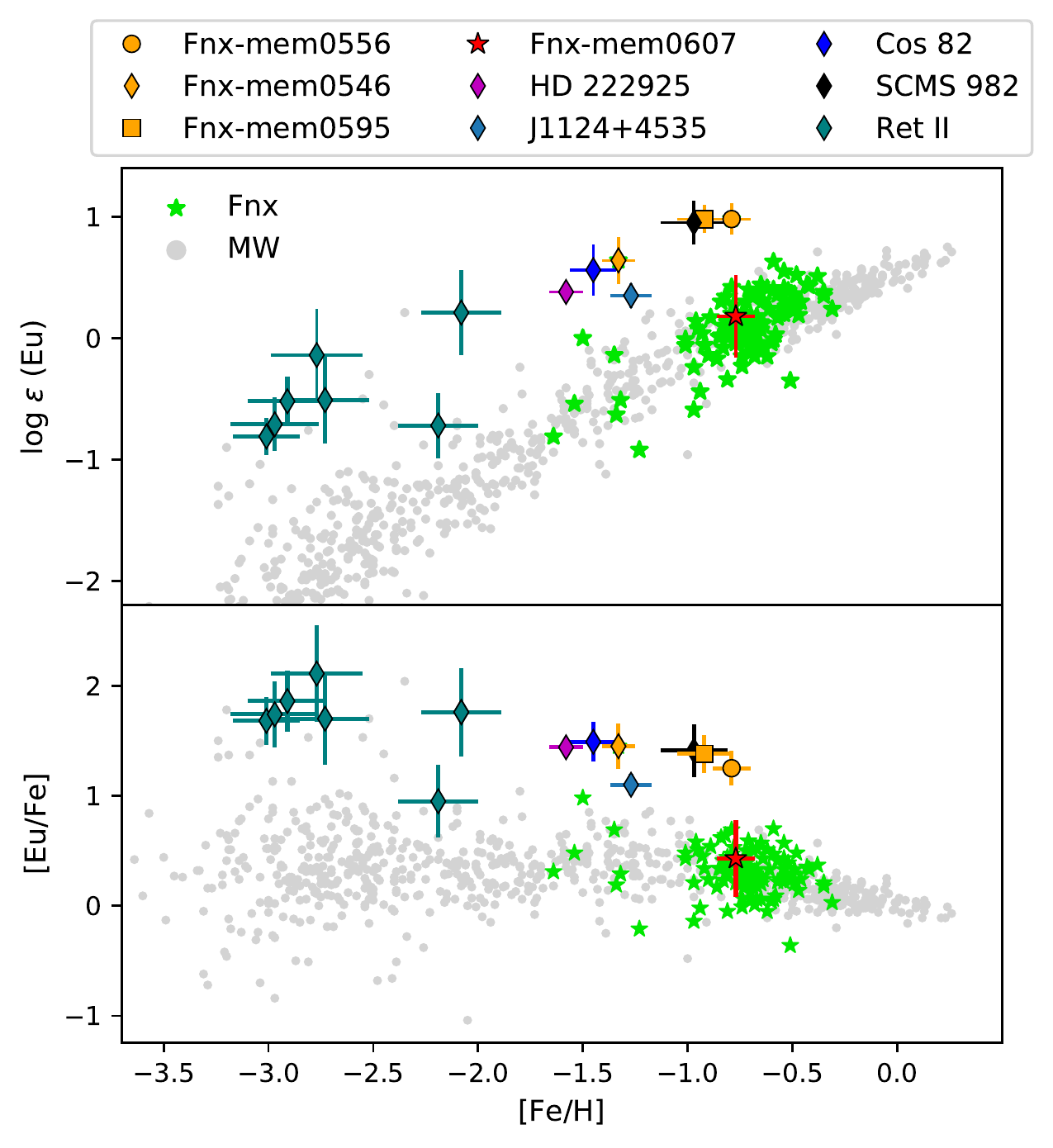}
      \caption{Absolute Eu (upper panel) and relative [Eu/Fe] abundances (lower panel) versus metallicity following the same notation and references as in Fig.~\ref{fig:mg_evolution}.} 
         \label{fig:eu_enhancement}
   \end{figure}

The three  Eu-enhanced Fornax stars  are enriched in europium with $\sim 1$ dex more than the average Fornax stars (see Fig.~\ref{fig:eu_enhancement}). With $\log \epsilon (\mathrm{Eu}) =0.98$, Fnx-mem0595 and Fnx-mem0556 are the most europium-enriched stars observed until today, together with  SCMS~982 that is enriched by an s- or i- process \citep[see][for a discussion]{Skuladottir2020b}. Even though rare, there are several stars with enhanced europium at low metallicities. However, such stars are even more rare at high metallicities.  In Fig.~\ref{fig:eu_enhancement}, we show some of those Eu-stars: Cos~82 \citep{Shetrone2001,Aoki2007,Sadakane2004}, J1124+4535 \citep{Xing2019}, HD~222925 \citep{Roederer2018}, and SCMS~982 \citep{Geisler2005,Skuladottir2020b}. With the exception of SCMS~982, these stars  show a dominant contribution from the r-process as indicated by their $\mathrm{[Ba/Eu]}$ ratio and  discussed below (Fig.~\ref{fig:ba_eu_theory}).
 
\begin{figure*}
   \centering
   \includegraphics[width=\hsize]{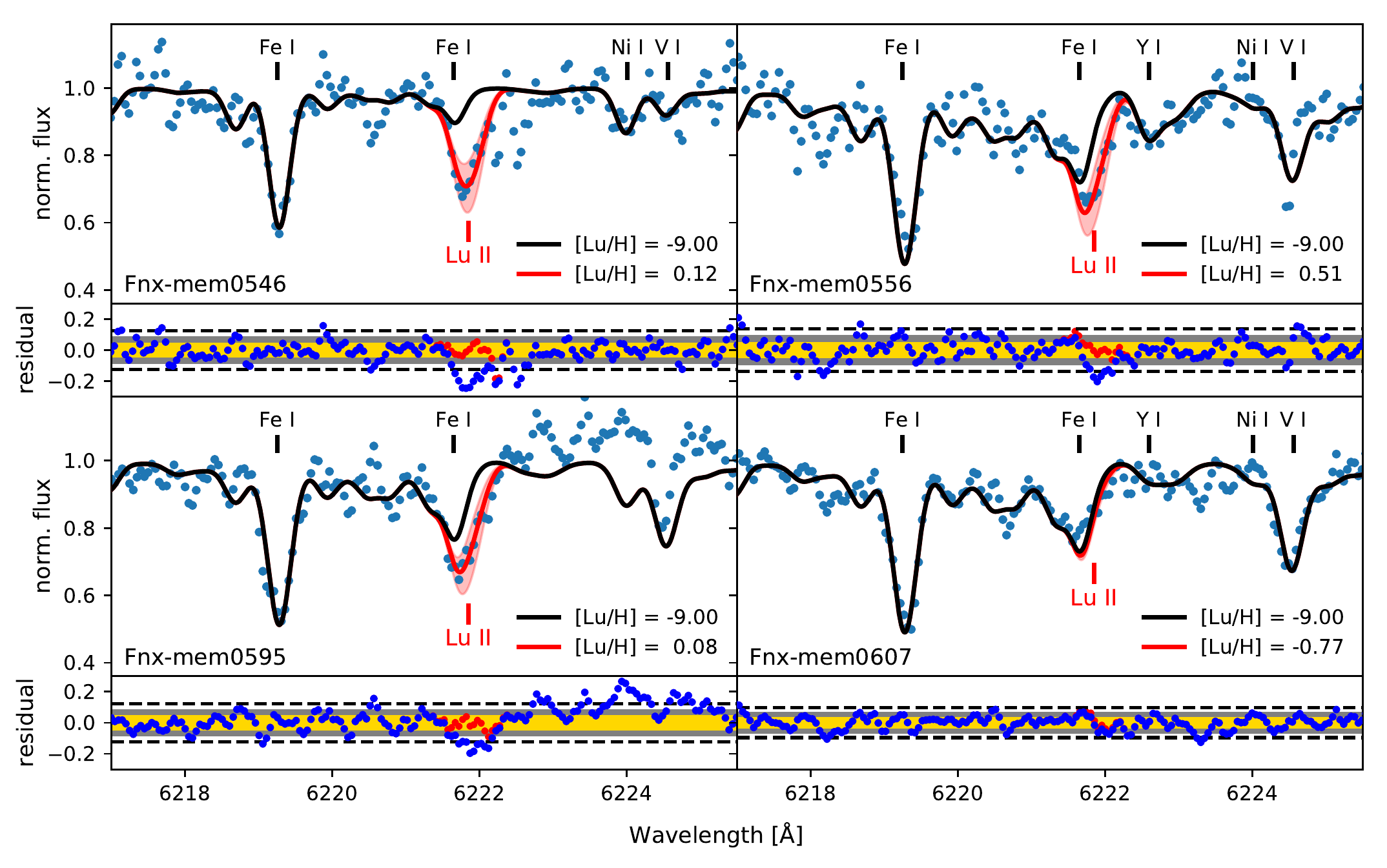}
      \caption{Normalized flux together with a synthetic spectrum. The synthetic spectrum is shown for a negligible amount of Lu (black line) and for an enhanced Lu abundance (red line). A red band indicates a deviation of the Lu abundance in the range of a typical error ($\pm 0.3$ dex). The lower panels show the residual of the synthetic spectrum with respect to the observed flux. Yellow horizontal bands mark a one sigma deviation from the synthesized spectra, grey bands a two sigma deviation, and dashed lines a three sigma deviation.} 
         \label{fig:lu_synth}
   \end{figure*}

The combination of metal-rich stars together with the extreme enhancements of neutron capture elements gives us the unique possibility to derive abundances of neutron capture elements that are otherwise challenging to detect. Therefore, we are able to derive lutetium abundances (Fig.~\ref{fig:lu_synth}).
   
Due to the heavy blending of the absorption line, we claim a Lu detection if more than three points deviate by more than three sigma (dashed line in Fig.~\ref{fig:lu_synth}) from the flux when assuming basically no Lu (i.e., $\mathrm{[Lu/H]} = -9$). Even though uncertain due to irregular noise next to the absorption line, we clearly detect lutetium in Fnx-mem0546, but also in Fnx-mem0556 and Fnx-mem0595. Because the continuum shows some artifacts next to the absorption line in Fnx-mem0595, we assign an additional error of $0.1\,\mathrm{dex}$ to the lutetium abundance. This adds all three stars to the rare stars with lutetium detections ($9$ in JinaBase, \citealt{Abohalima2018}, $13$ in SAGA, \citealt{SAGA}) and makes it the first detections in a dwarf spheroidal galaxy.
\newpage
\section{Origin of Eu-stars}
\label{sec:origin}

In this section we probe the origin of the Eu-stars by first finding the predominant process that enriched these stars. Following, we assess which possible sites could host this process. We make use of yield predictions and abundance ratios for this purpose, but we caution the use of a single element ratio to assign a dominant process contribution. We point towards using such ratios together with observationally derived abundance patterns, as these present a more complete chemical trace of the true stellar enrichment.

\subsection{r-, i-, or s-process?}
\label{sec:pattern}

\begin{figure*}
   \centering
   \includegraphics[width=\hsize]{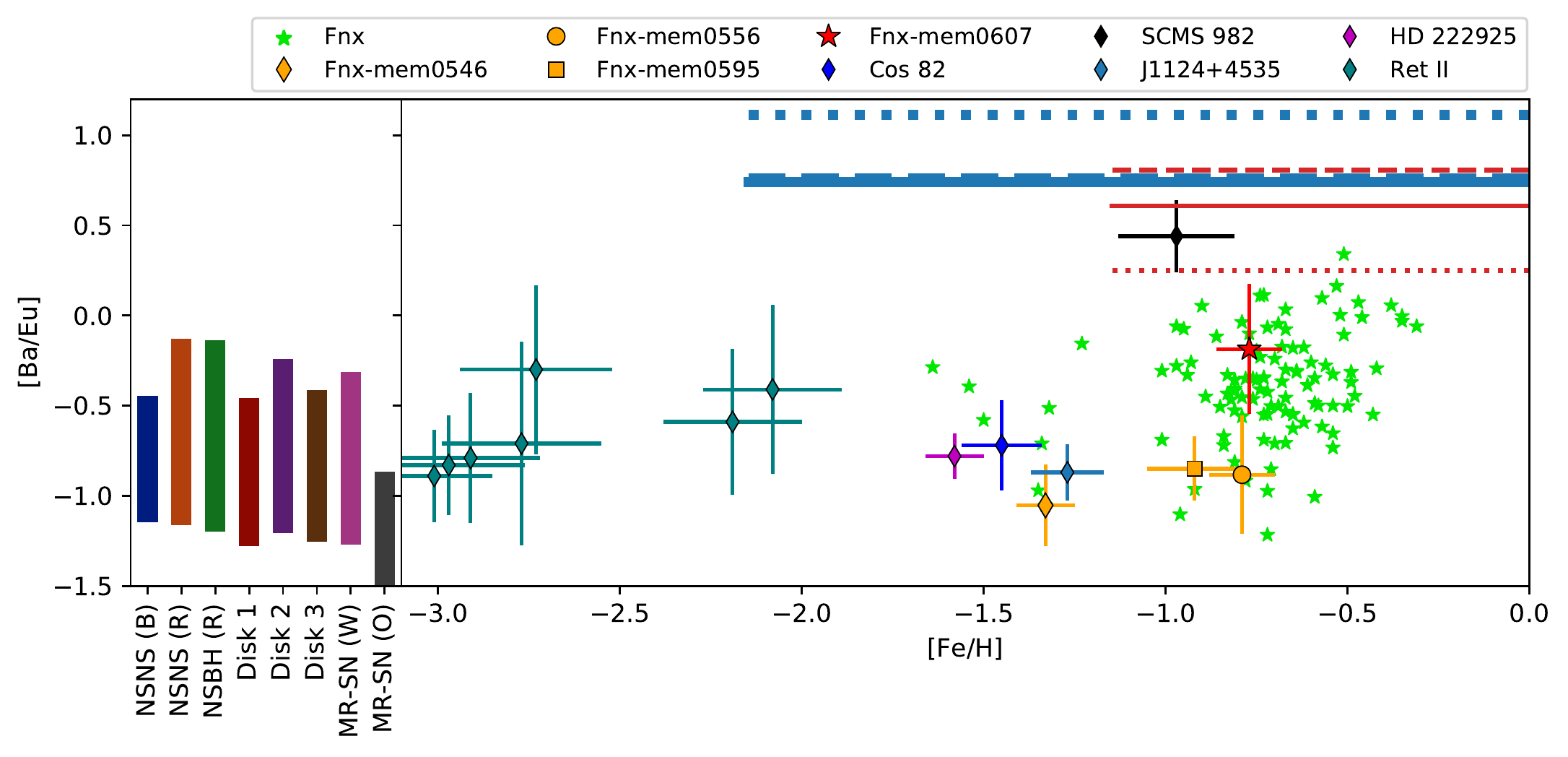}
      \caption{[Ba/Eu] versus metallicity (right panel) following the same notation and references as in Fig.~\ref{fig:mg_evolution}.  Theoretical predictions are shown for  s-process (horizontal lines) and r-process (left panel). The s-process yields are from F.R.U.I.T.Y. database \citep{Cristallo2011} and are indicated by horizontal lines starting at the metallicity of the model. These models are for three AGB masses: $1.3\,M_\odot$ (solid lines), $2.0\,M_\odot$ (dashed lines), and $5.0\,M_\odot$ (dotted lines). The red, thin lines indicate a metallicity of $Z=0.001$, and blue, thick lines correspond to $Z=0.0001$ and [$\alpha$,Fe]=0.5. The r-process ratios are shown with bars that cover the calculate range taking into account uncertainties due to the various astrophysical conditions and different nuclear physics theoretical models. The ratio is shown for the early, dynamical ejecta of neutron star merger (NSNS) and neutron star black hole merger (NSBH) as well as for their disk ejecta (Disk~1, Disk~2, and Disk~3). Results are also presented for MR-SNe based on two simulations. For more details and references see Sect.~\ref{sec:chasing} and \cite{Cote2020, Eichler2019, Reichert2021}.} 
         \label{fig:ba_eu_theory}
\end{figure*}

At high metallicity, the neutron-capture elements stem mainly from  the s- and r-process, possibly also from the i-process. The i-process operates at intermediate n-densities and exposures and may be able to take place in rapidly accreting white dwarfs \citep[e.g.,][]{Denissenkov2017}, super asymptotic giant branch stars \citep[e.g.,][]{Doherty2015}, or low-mass AGB \citep[e.g.,][]{Stancliffe2011}. In order to check whether the Eu-stars are enriched by an r-process event or whether they have a strong contribution from the s-process, we first check the [Ba/Eu] ratio shown in Fig.~\ref{fig:ba_eu_theory} and Table~\ref{tab:abundances}. For the three stars, the ratio is well below zero and this is often used to define neutron-capture-rich, r-process enhanced stars ([Ba/Eu]$<$0, \citealt{Beers2005}). A stronger constraint on the heavy-element enhancement is the estimate of the pure process trace, which is typically assessed through the [Ba/Eu] being less than $\sim-0.7$\,dex (pure r) or larger than $\sim1.1$\,dex  (pure s) \citep{Arlandini1999,Hansen2018}. The [Ba/Eu] ratio points towards a pure r-ratio in the three Fornax stars, while the reference star, Fnx-mem0607, is clearly a mixture of r and s. 

We also probe if there is an i-process contamination. A typical fingerprint of the i-process is a positive [Ba/La] ratio \citep{Koch2019,Hampel2016,Denissenkov2019}, but this ratio is negative for the four Fornax stars. In this connection, we note that Fnx-mem0546 has a relatively low Ba abundance with respect to La and also compared to the Ba in the other two stars. The La abundance is more than 1\,dex higher than the Ba abundance in Fnx-mem0546. This leads to a strong odd-even abundance difference in this region. In Fig.~\ref{fig:ba_eu_theory}, it is also clear that SCMS~982 is not r-process dominated with [Ba/Eu] close to 0.5.  If the peculiar pattern of SCMS~982 stems from an s- or i-process is still under debate \citep{Geisler2005,Skuladottir2020b}. Based on these abundance ratios, we focus our further comparison on r-process and s-process yields.

We now compare the observationally derived abundances to theoretical predictions. For the s-process, we take yields for different AGB star masses and metallicities from the F.R.U.I.T.Y. database \citep{Cristallo2011,Cristallo2015}. In Fig.~\ref{fig:ba_eu_theory}, this contribution is shown by horizontal dashed and dotted lines covering the metallicity range where they are predicted and appropriate. For the r-process, we compare to yields from different scenarios (for details see Sect.~\ref{sec:chasing}). The variation in any abundance ratio within a given scenario is due to the  nuclear physics uncertainties \cite[see][for more details about models and nuclear physics included in the nucleosynthesis]{Cote2020,Eichler2019}. As seen from Fig.~\ref{fig:ba_eu_theory}, the [Ba/Eu] ratios of the three Eu-stars agree with the theoretical predictions of the r-process within uncertainties. However, those uncertainties are rather large and thus it is not possible to conclude which scenario contributed most to the observed abundances.

\begin{figure}[h]
   \centering
   \includegraphics[width=\hsize]{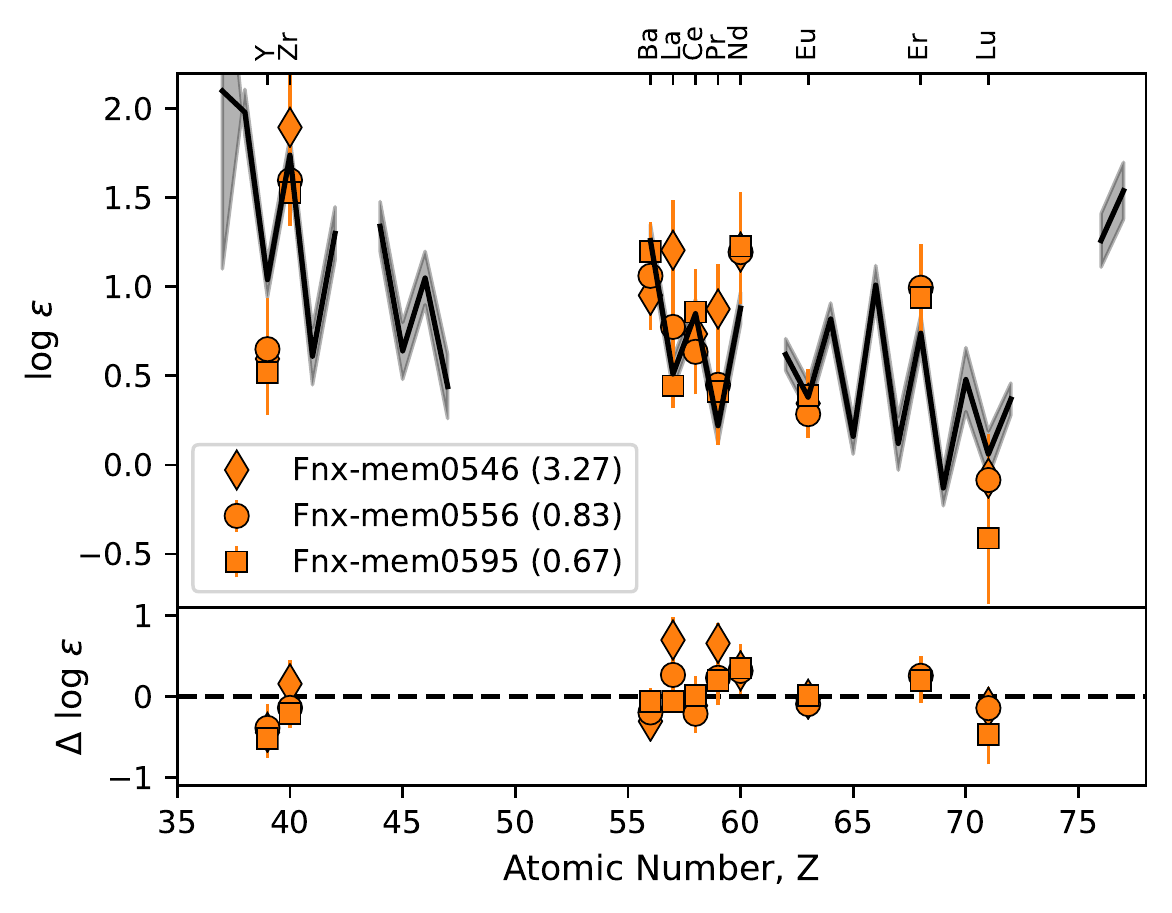}
      \caption{Abundances of Fnx-mem0546 (diamonds), Fnx-mem0556 (circles), and Fnx-mem0595 (squares) scaled to fit the r-process pattern of  HD~222925 \citep[black line, with errors indicated as grey bands, ][]{Roederer2018}. In the legend we give the reduced $\chi ^2$ of the fit for  $Z>50$. Bottom panel shows the relative differences between the Fornax stars and HD~222925.} 
         \label{fig:r_pattern}
   \end{figure}

Now we explore another possibility to check whether the three Eu-star's abundances are mainly due to the r-process. We compare their abundance patterns to that of HD 222925 \citep{Roederer2018}, a star that has been identified as r-process enhanced. Figure~\ref{fig:r_pattern} shows this comparison and demonstrates that two of the Eu-stars (Fnx-mem0556 and Fnx-mem0595) fit with an excellent reduced $\chi^2<1$  the r-process pattern of HD 222925. Fnx-mem0546 deviates slightly from this pattern with an enhancement in lanthanum, and praseodymium relative to barium and europium. Lutetium on the other hand agrees well with the r-process pattern for Fnx-mem0546 and Fnx-mem0556. For Fnx-mem0595, lutetium is underabundant compared to the r-process ratio, however, it also has a larger error (see Sect.~\ref{sec:heavy_ncap}).
In summary, all stars show a striking similarity to the r-process pattern without too many contributions from other processes. We can therefore safely assume that the abundances of the three Eu-stars are mainly enriched by the r-process.

\subsection{One r-process event}
\label{sec:one-r}

After showing that the Eu-stars are r-process enhanced, we estimate how many r-process events are necessary to enrich these stars and calculate the ejected europium per event. The three Eu-stars were born in a gas cloud/region with a total mass $M_\mathrm{gas}$.  \cite{Magg2020} provide an approximation for the dilution mass where the new elements from an r-process event(s) are mixed:
\begin{equation}
\label{eq:dillmass} 
M_\mathrm{gas} = 1.9 \cdot 10^4 \mathrm{M}_\odot E_{51}^{0.96}n_0^{-0.11}, 
\end{equation}
here $n_0$ is the ambient density and  $E_{51}$ is the explosion energy of the event. \cite{Magg2020} uses  $n_0 = 1$ corresponding to the environment where Pop III stars formed, however this has an small impact on the final dilution mass due to the small exponent. The explosion energy ($E_{51}$) of the  event can be assumed to be between $10^{51}$ and  $10^{52} \, \mathrm{erg}$ (c.f., $10^{51}$ erg for GW170817; \citealt{Kathirgamaraju2019} and the explosion energy of $10^{52}$ erg of hypernovae; \citealt{Nomoto2006}). The dilution mass ($M_\mathrm{gas}$) therefore lies between $10^4$ to $10^5\, \mathrm{M_\odot}$, which is at least three orders of magnitudes smaller than the total stellar mass of Fornax $M_\mathrm{Fornax}\sim10^8\, \mathrm{M_\odot}$ \citep{Lokas2009}. 

Before the r-process event(s), the mass of Eu in the gas was $M_\mathrm{Eu}^\mathrm{pre-event}$ and the mass injected by the event was $M_\mathrm{Eu}^\mathrm{r-event}$. We can estimate $M_\mathrm{Eu}^\mathrm{pre-event}$ using the abundance of our reference star, Fnx-mem0607, in the following definition:
\begin{equation}
    \log \epsilon (\mathrm{Eu}) = \log \frac{M_\mathrm{Eu}^\mathrm{pre-event}}{A_\mathrm{Eu} \cdot M_\mathrm{gas}} +12,
\end{equation}
where $A_\mathrm{Eu}=152$ is the average mass number of europium. Therefore, the Eu  in the r-process stars is:
\begin{equation}
    \log \epsilon (\mathrm{Eu}) = \log \frac{M_\mathrm{Eu}^\mathrm{pre-event}+M_\mathrm{Eu}^\mathrm{r-event}}{A_\mathrm{Eu} \cdot M_\mathrm{gas}} +12,
    \label{eq:Eu_event}
\end{equation}
and from this expression we  obtain an europium mass per event $M_\mathrm{Eu}^\mathrm{r-event}$ between \mbox{$\sim 1.5\cdot 10^{-5}$} and \mbox{$\sim 3\cdot 10^{-4} \, \mathrm{M_\odot}$}. This is similar to the values reported by \citet{Ji2016a} to explain the r-process enhanced stars in Reticulum II \mbox{($\sim 2.5\cdot 10^{-5} - 5\cdot 10^{-5} \, \mathrm{M_\odot}$)}. Therefore, it is  likely that Fnx-mem0556, Fnx-mem0595, and Fnx-mem0546 also got enriched by a single r-process event. Moreover, the stars probably  formed only a few Myr after the event. Otherwise, the r-process material would have been mixed  into a larger amount of gas \citep{vandefoort2019}.

\begin{figure}
  \centering
  \includegraphics[width=\hsize]{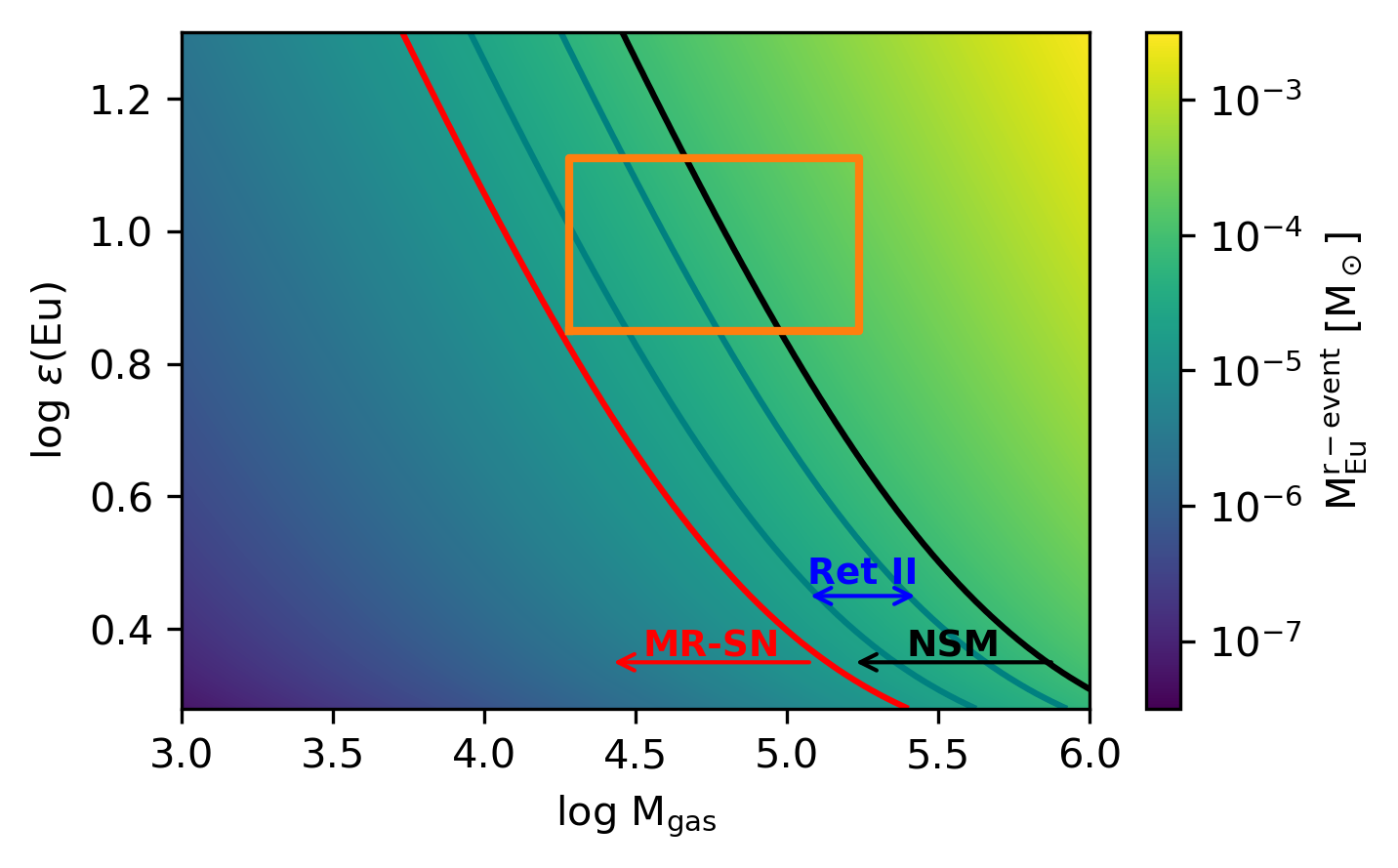}
     \caption{Absolute abundances of europium versus dilution mass. The amount of europium needed to enhance a star to a given absolute abundance value is color coded. The orange box indicates the range, predicted from Fnx-mem0556, where the left and right vertical lines are obtained from different explosion energies and the horizontal lines from the observed absolute europium abundances including the  error. The range of estimated Eu masses to enhance Reticulum II \citep{Ji2016a} is given by blue lines. The maximum Eu mass ejected  from NSMs based on simulation is shown as a black line ($8\cdot 10^{-5}\,\mathrm{M_\odot}$), while the maximum ejected europium mass from MR-SNe simulations is traced by red line ($1.5\cdot 10^{-5}\,\mathrm{M_\odot}$).} 
        \label{fig:ejected_eu}
 \end{figure}

Our simple estimate has several input parameters (e.g., $n_0$, $E_{51}$), therefore we show in Fig.~\ref{fig:ejected_eu} an overview covering different possible values. In this figure, the absolute europium abundances of Fnx-mem0556 are indicated by horizontal orange lines, the estimated dilution mass by vertical orange lines, and the colors correspond to the Eu mass ejected by the r-process event using Eq.~\ref{eq:Eu_event}. The latter changes smoothly over several orders. However, the amount of Eu needed to explain the Eu-stars ($M_\mathrm{Eu}\sim 10^{-5}-10^{-4}\, M_\odot$, Fig.~\ref{fig:ejected_eu}) agrees with having only one r-process event (c.f. also \citealt{Beniamini2016} for estimates of the ejected Eu mass per r-event in dSph galaxies.).

 \subsection{Chasing the r-process site in Fornax}
\label{sec:chasing}

Which r-process event is responsible for the enhanced Eu in the three Fornax stars? The possible candidates are compact binary mergers (two neutron stars or a neutron star and a black hole) and MR-SNe including the early explosion phase and the late evolution as collapsar.

In Fig.~\ref{fig:ba_eu_theory}, we have compared the [Ba/Eu] of the Eu-stars to various theoretical predictions from compact binary mergers (dynamical and disk ejecta) and from MR-SNe using different nuclear physics input \citep{Kodama1975,Duflo1995,Panov2001,Moeller2003,Kelic2008,Marketin2016}. We test the dynamical ejecta of mergers of two neutron stars (NSNS (B), \citealt{Bovard2017}) and (NSNS (R), \citealt{Korobkin2012}), and a neutron star and a black hole (NSBH (R), \citealt{Korobkin2012}); ejecta from the accretion disk formed after merger surrounding the compact object (Disk 1, Disk 2, and Disk 3, \citealt{Wu2016}); ejecta from magneto-rotational supernovae MR-SN (W) \citep{Winteler2012} and MR-SN (O) \citep{Reichert2021}. A similar set of models was used also by \citet{Cote2020,Eichler_Arcones2021} to compare to meteorite ratios of r-process isotopes. However, there are too large theoretical uncertainties in the nuclear physics input as well as the variation in the astrophysical conditions and this makes it impossible to conclude from the [Ba/Eu] ratio alone which scenario has contributed to the enhanced Eu. 

The mass of Eu estimated above ($M_\mathrm{Eu}\sim 10^{-5} - 3\cdot 10^{-4} \, \mathrm{M_\odot}$) could be also used as a constraint on the r-process site. We compare this amount of Eu to the mass that is ejected in NSM and MR-SNe, as obtained in hydrodynamic simulations. However, these simulations are still uncertain due to various  aspects (e.g., resolution, magnetic fields, neutrino matter interactions, high density equation of state) and predictions about the amount of Eu  ejected are only approximate.

In compact binary mergers, the masses of ejected europium in the dynamic ejecta depends on the simulation and can differ by orders of magnitudes (see, e.g., \citealt{Cote2018} for an overview of different ejecta masses of the dynamical ejecta). Just to name a few examples, the europium mass based on the simulation of \citet{Bovard2017} is $M_\mathrm{Eu} \sim 10^{-6}-10^{-5} \, \mathrm{M_\odot}$. Yields of other simulations are slightly higher with $M_\mathrm{Eu} \sim 10^{-5}-8 \cdot 10^{-5} \, \mathrm{M_\odot}$ \citep{Korobkin2012} and similar $M_\mathrm{Eu} \sim 3\cdot 10^{-5}-5 \cdot 10^{-5} \, \mathrm{M_\odot}$ \citep{Goriely2011}. In the case of a NSBH merger, the ejected europium mass of the dynamic ejecta can reach $M_\mathrm{Eu} \sim 3\cdot 10^{-4}\,\mathrm{M}_\odot$ \citep{Korobkin2012}. In addition, the disk ejecta will also contribute to  europium ($M_\mathrm{Eu} \sim 3\cdot 10^{-6}-  10^{-5} \, \mathrm{M_\odot}$, \citealt{Wu2016}). However, it is still under discussion how much each ejected component contributes and how neutron-rich the components are \citep[see, e.g.,][for a recent review]{Shibata2019}.

In GW170817, \cite{Watson2019} directly observed Sr and predicted a mass  $M(\mathrm{Sr})\ga 5 \cdot 10^{-5}\, \mathrm{M_\odot}$ . When assuming that the ejecta is scaled like the solar r-process residual \citep{Sneden2008} the ejected europium mass is  $M_\mathrm{Eu} \sim 3\cdot 10^{-6}\,\mathrm{M}_\odot$. We note that the lighter heavy elements like strontium usually scatter with respect to the heavier elements. Because of this and other involved uncertainties, this value should be treated as a rough estimate only. Nevertheless, it agrees well with the masses of $M_\mathrm{Eu} \sim 3\cdot 10^{-6}-1.5 \cdot 10^{-5}\,\mathrm{M}_\odot$ obtained in \citet{Cote2018} for GW170817.

The amount of Eu produced in MR-SNe based on recent simulations is  $\lesssim 1\cdot 10^{-5}\, \mathrm{M_\odot}$ \citep{Winteler2012}, $\lesssim 1.5\cdot 10^{-5}\, \mathrm{M_\odot}$ \citep{Nishimura2015}, $\lesssim 8\cdot 10^{-6}\, \mathrm{M_\odot}$ \citep{Nishimura2017}, and $\lesssim 5\cdot 10^{-6}\, \mathrm{M_\odot}$ \citep{Reichert2021}. These values are on the lower end of our assumed uncertainties (Fig.~\ref{fig:ejected_eu}), but they could still be responsible for the enrichment. After some MR-SNe, a black hole forms  and is surrounded by an accretion disk. The disk conditions may be similar to those found in accretion disks after neutron star mergers and may favour an r-process. \cite{Siegel2019} have found that Eu may be produced in collapsars, however, more detailed investigations are necessary to understand whether the conditions are appropriate for an r-process, see \cite{Miller2019,Just2021}.

In Fig.~\ref{fig:ejected_eu}, we show the maximum  europium mass ejected by NSM (black line, $8\cdot 10^{-5}\,\mathrm{M_\odot}$) and by MR-SNe (red line, $1.5\cdot 10^{-5}\,\mathrm{M_\odot}$). Moreover, we include the estimated europium mass necessary to explain r-II stars  in Reticulum II \citep{Ji2016a}. All of these estimates are close to the mass that is necessary to enhance the Eu-stars, as derived from the observed stellar abundances.

In addition to Eu, one can use $\alpha-$elements, e.g., Mg, to check whether MR-SNe are the r-process site responsible for more element abundances in the Fornax Eu-stars. Since the amount of ejected $\alpha-$elements depends on the progenitor mass \citep[e.g.,][]{Kobayashi2006}, and the progenitors of MR-SNe may differ from the average CC-SNe, the amount of ejected $\alpha$-elements may also vary between the two types of SNe.
If an abnormal supernova produces r-process material and a huge amount of Mg, one would possibly see a signature of this in the Mg abundances in the Fornax Eu-stars. However, in Fig.~\ref{fig:mg_evolution}, the three stars present normal Mg abundances (see also Table~\ref{tab:abundances}). This may be an indication of a neutron star merger producing the observed Eu or of MR-SNe ejecting normal amount of Mg as any other supernova in Fornax. Therefore, we cannot use Mg as an indicator for rare supernovae and their r-process yields.

In summary, based on the abundance ratio of [Ba/Eu], on the estimated amount of Eu, and on the Mg abundance, we cannot determine the r-process site enriching the Eu-stars in Fnx. Based on Fig.~\ref{fig:ejected_eu}, NSMs seem a promising site to explain the r-material in the Eu-stars, however, their time delay coupled with recent star formation in Fnx may pose an issue that is easier to overcome if MR-SNe would be the r-site. On the other hand, recent studies also show that a substantial fraction of the NSM population may occur on short timescales ($< 100\, \mathrm{Myr}$, see e.g. \citealt{Beniamini2019}).

\subsection{Time scale - late r-process event linked to a star formation burst in Fornax}
\label{sec:environment}

The strong r-process enhancement, despite the remaining unknown formation site, combined with the typical low $\alpha-$ content indicates, together with the \textit{Gaia} proper motions \citep[DR2,][]{Gaia2018,Gaia_vizier2018}, that these rare Eu-stars are indeed Fornax members (see Sect.~\ref{sec:sample}).

From the time scale perspective, neutron star mergers remain viable sources and could explain the observed abundances. Neutron star mergers are expected to contribute with some delay and thus such an event could account for producing r-process material late or at high metallicities. However, this would imply that there is some star formation just after the neutron stars merge and eject r-process material. Similarly linked to the late star formation, massive stars could also have formed and exploded as MR-SNe producing the r-process observed in the Eu-stars. This would require more gas at a later stage, which may be feasible in Fornax. In any case, the three r-rich stars must have formed shortly after the r-process event, i.e., there must have been an active star forming environment. According to \citet{Lemasle2014}, Fnx-mem0556 has an age of $4.36\pm0.86\, \mathrm{Gyr}$ and Fnx-mem0595 an age of $5.75\pm 1.78 \, \mathrm{Gyr}$. Fornax shows a complex star formation history with several outbreaks. A significant number of stars were formed at early times, i.e., more than $10\, \mathrm{Gyr}$ ago.  Moreover, there was a very recent period of star formation $\sim 4 \, \mathrm{Gyr}$ ago \citep[e.g.,][]{Coleman2008,deBoer2012,Hendricks2014,Weisz2014,Rusakov2021}. This late sudden burst of star formation  agrees well with the age of the r-process enhanced stars. Although the origin of this burst is not known, there are speculations that Fornax has undergone some galaxy merger events in its history, indicated by over-dense features in the spatial distribution of stars in Fornax \citep[e.g.,][]{Coleman2004}.

Another explanation for such an peculiar enhancement of neutron-capture elements was investigated by \citet{Tsujimoto2002}. They proposed that inhomogeneous mixing, caused by a high velocity dispersion in dSph galaxies may be responsible for the existence of stars as Cos~82, and possibly also Fnx-mem0546, Fnx-mem0556, and Fnx-mem0595 \citep[see also][for a discussion]{Sadakane2004,Aoki2007}. 
\section{Conclusions}
\label{sec:conclusions}

We study neutron-capture elements in Fornax stars including three stars at high metallicities with extreme enhancements of heavy r-process elements. We define these new type of stars as Eu-stars, they are r-II stars ([Eu/Fe]$>1$) at high metallicity ([Fe/H]$\gtrsim -1.5$). Although previously observed, metal-rich stars with enhanced neutron-capture elements are rare. The Eu-stars show an europium abundance up to $\log \epsilon (\mathrm{Eu})= 0.98 \pm 0.12$, which is  the highest europium abundance ever observed. Thus, Eu is approximately three times more abundant than in our Sun while their iron abundance is a factor of seven smaller. In addition to an enrichment in heavy neutron-capture elements,  Zr is also enhanced in the three stars. In order to compare to other Fornax stars, we have derived Zr abundances for $105$ stars. This is the largest Zr sample in a dSph to date and offers a new chance to explore the galactic chemical evolution of Fornax. Moreover,  we have derived lutetium abundances for  the first time for stars in a dSph galaxy. This was possible due to the unique combination of high metallicities and r-process enrichment that allowed for a Lu line detection.

We have demonstrated that the enhancements in neutron-capture elements is due to the r-process as indicated by the [Ba/Eu] ratio and typical r-process pattern. Moreover, we give an estimate of the amount of Eu necessary to explain these r-process rich / Eu-stars, namely  $M(\mathrm{Eu)}\sim8\cdot 10^{-6} - 3\cdot 10^{-4} \, \mathrm{M_\odot}$. This agrees with an individual r-process event being enough to explain the observed Eu abundances. Based on the elemental ratios and the  europium mass ejected, we try to identify the r-process site by comparing to theoretical yield predictions from neutron star mergers and magneto-rotational supernovae. However, the uncertainties in the astrophysical conditions and the nuclear physics input prevent us of making any firm conclusion about the site. There is a clear need of improved hydrodynamic simulations with detailed microphysics as well as more theoretical and experimental information of the extreme neutron-rich nuclei involved in the r-process.

The r-process event responsible for the Eu-stars was occurring during a star formation episode. If the event was a NSM, this could come from neutron stars born in early supernovae. The delay of the merger was coinciding with the star formation event where the Eu-stars were born shortly after. Despite NSM yield a very promising range of Eu abundances, the time scale is a bit more tricky in this scenario. Another possibility is that during the star formation event, massive stars formed and at least one died fast as a MR-SN ejecting the r-process material necessary to explain the observed abundances. Therefore, an active star forming environment simultaneously with the r-process event is  in any case required for the formation of stars with such an enhanced europium content. The age of the stars approximately traces the time when the r-process event occurred. Their age is estimated to be around $4\, \mathrm{Gyr}$ \citep{Lemasle2014}, which coincides with a sudden increase of star formation in Fornax \citep{Coleman2008,deBoer2012,Hendricks2014,Weisz2014,Rusakov2021}. 
We conclude that the existence of these Eu-stars proves that the r-process  can efficiently form r-II  stars across a broad range of dwarf galaxies - from the faintest low-mass ones to the most massive dSph galaxies. It is not unique to ultra faint dwarf galaxies as suggested before. Moreover, we  emphasize that  gas dilution and star formation time scales must be considered in the search for the r-process sites. Future observations are critical to find more Eu-stars that are key to understand the origin of heavy elements produced by the r-process.
\acknowledgments
The authors thank M. Eichler, M. Hanke, A. Koch, and {\'A}. Sk{\'u}lad{\'o}ttir for valuable discussions. 
MR and AA were supported by the ERC Starting Grant EUROPIUM-677912, Deutsche Forschungsgemeinschaft through SFB~1245, and Helmholtz Forschungsakademie Hessen für FAIR. CJH acknowledges support from the Max Planck Society. 
This work has benefited from the COST Action “ChETEC” (CA16117) supported by COST (European Cooperation in Science and Technology).

\appendix

\section{Zirconium abundances}
We list all derived zirconium abundances together with metallicities from \citet{Reichert2020} in Table~\ref{tab:abundances_zr}.
\begin{table*}
\caption{Metallicities and zirconium abundances of $105$ stars in Fornax.}  
\vspace*{-0.35cm}
\label{tab:abundances_zr}     
\centering                        
\begin{tabular}{l c c | l c c}      
\hline\hline               
ID & {[Fe/H]} & $\log \epsilon (\mathrm{Zr})$ & ID & {[Fe/H]} & $\log \epsilon (\mathrm{Zr})$\\   
\hline                      
{[LHT2010] BL147} & $-1.50 \pm 0.05$ & $1.81 \pm 0.22$            & {[WMO2009] For-0956} & $-0.72 \pm 0.08$ & $1.81 \pm 0.17$\\
{[WMO2006] F01-20} & $-1.36 \pm 0.05$ & $1.41 \pm 0.11$           & {2MASS J02384113-3444205} & $-0.71 \pm 0.05$ & $1.57 \pm 0.13$\\
{[LDH2014] Fnx-mem0546} & $-1.33 \pm 0.08$ & $2.35 \pm 0.27$      & {2MASS J02401677-3429346} & $-0.71 \pm 0.07$ & $1.80 \pm 0.13$\\
{[WMO2009] For-0970} & $-1.23 \pm 0.05$ & $1.39 \pm 0.11$         & {[MOW91]   8} & $-0.70 \pm 0.10$ & $1.75 \pm 0.13$\\
{[WMO2009] For-0361} & $-1.01 \pm 0.11$ & $1.57 \pm 0.19$         & {[LDH2014] Fnx-mem0715} & $-0.70 \pm 0.13$ & $1.77 \pm 0.18$\\
{2MASS J02401043-3425177} & $-1.01 \pm 0.10$ & $1.44 \pm 0.10$    & {[WMO2009] For-0391} & $-0.69 \pm 0.07$ & $1.90 \pm 0.15$\\
{[WMO2009] For-1877} & $-0.97 \pm 0.09$ & $1.53 \pm 0.13$         & {2MASS J02383503-3441380} & $-0.69 \pm 0.06$ & $1.83 \pm 0.22$\\
{[LDH2014] Fnx-rgb0553} & $-0.97 \pm 0.05$ & $1.64 \pm 0.14$      & {2MASS J02390157-3436488} & $-0.68 \pm 0.08$ & $1.68 \pm 0.12$\\
{[WMO2009] For-0968} & $-0.96 \pm 0.06$ & $1.61 \pm 0.20$         & {2MASS J02395144-3421211} & $-0.68 \pm 0.09$ & $1.84 \pm 0.22$\\
{[MOW91]  25} & $-0.95 \pm 0.09$ & $1.60 \pm 0.16$                & {[LDH2014] Fnx-mem0574} & $-0.67 \pm 0.05$ & $1.90 \pm 0.18$\\
{[LHT2010] BL100} & $-0.94 \pm 0.10$ & $1.55 \pm 0.14$            & {2MASS J02391102-3428348} & $-0.67 \pm 0.10$ & $1.86 \pm 0.15$\\
{GB07 Fnx11} & $-0.93 \pm 0.06$ & $1.83 \pm 0.14$                 & {2MASS J02390031-3430302} & $-0.67 \pm 0.05$ & $1.62 \pm 0.14$\\
{[LDH2014] Fnx-mem0595} & $-0.92 \pm 0.13$ & $2.10 \pm 0.15$      & {[LDH2014] Fnx-mem0638} & $-0.67 \pm 0.07$ & $1.60 \pm 0.15$\\
{[LDH2014] Fnx-mem0675} & $-0.92 \pm 0.11$ & $1.83 \pm 0.19$      & {[LDH2014] Fnx-mem0631} & $-0.67 \pm 0.08$ & $1.76 \pm 0.13$\\
{[WMO2009] For-1579} & $-0.90 \pm 0.06$ & $1.80 \pm 0.14$         & {2MASS J02401790-3427010} & $-0.67 \pm 0.07$ & $1.93 \pm 0.18$\\
{2MASS J02392805-3434013} & $-0.89 \pm 0.08$ & $1.41 \pm 0.13$    & {[WMO2009] For-0365} & $-0.65 \pm 0.06$ & $1.78 \pm 0.12$\\
{[LDH2014] Fnx-mem0629} & $-0.88 \pm 0.05$ & $1.68 \pm 0.20$      & {2MASS J02401752-3426065} & $-0.65 \pm 0.09$ & $1.89 \pm 0.15$\\
{2MASS J02385721-3435400} & $-0.86 \pm 0.08$ & $1.72 \pm 0.15$    & {[KGS2010] For   64059} & $-0.65 \pm 0.10$ & $1.82 \pm 0.20$\\
{2MASS J02392022-3431571} & $-0.85 \pm 0.07$ & $1.72 \pm 0.14$    & {[WMO2009] For-0387} & $-0.65 \pm 0.11$ & $1.93 \pm 0.14$\\
{[WMO2009] For-1219} & $-0.84 \pm 0.06$ & $1.78 \pm 0.14$         & {2MASS J02385365-3433048} & $-0.64 \pm 0.08$ & $1.81 \pm 0.13$\\
{2MASS J02394528-3431581} & $-0.84 \pm 0.04$ & $1.66 \pm 0.17$    & {[WMO2009] For-0910} & $-0.64 \pm 0.05$ & $2.00 \pm 0.22$\\
{[LDH2014] Fnx-mem0626} & $-0.83 \pm 0.07$ & $1.64 \pm 0.11$      & {2MASS J02391606-3430135} & $-0.62 \pm 0.08$ & $2.01 \pm 0.12$\\
{2MASS J02393153-3423052} & $-0.83 \pm 0.11$ & $1.66 \pm 0.14$    & {2MASS J02390853-3430556} & $-0.62 \pm 0.07$ & $1.95 \pm 0.12$\\
{[LHT2010] BL158} & $-0.82 \pm 0.07$ & $1.81 \pm 0.19$            & {2MASS J02395427-3435114} & $-0.61 \pm 0.11$ & $1.94 \pm 0.20$\\
{2MASS J02383502-3442406} & $-0.81 \pm 0.06$ & $1.86 \pm 0.22$    & {2MASS J02391398-3428364} & $-0.60 \pm 0.11$ & $1.81 \pm 0.12$\\
{GB07 Fnx08} & $-0.81 \pm 0.05$ & $1.74 \pm 0.16$                 & {[LDH2014] Fnx-mem0682} & $-0.59 \pm 0.06$ & $1.89 \pm 0.12$\\
{[LHT2010] BL204} & $-0.81 \pm 0.03$ & $1.69 \pm 0.12$            & {[LDH2014] Fnx-mem0717} & $-0.59 \pm 0.08$ & $1.78 \pm 0.18$\\
{[LHT2010] BL261} & $-0.81 \pm 0.11$ & $1.51 \pm 0.17$            & {2MASS J02391437-3434427} & $-0.59 \pm 0.09$ & $2.09 \pm 0.17$\\
{[LDH2014] Fnx-mem0633} & $-0.81 \pm 0.12$ & $1.65 \pm 0.20$      & {[LDH2014] Fnx-mem0678} & $-0.58 \pm 0.07$ & $1.93 \pm 0.22$\\
{2MASS J02391159-3430448} & $-0.79 \pm 0.07$ & $1.68 \pm 0.12$    & {2MASS J02391783-3430570} & $-0.57 \pm 0.10$ & $1.90 \pm 0.15$\\
{[LDH2014] Fnx-mem0634} & $-0.79 \pm 0.08$ & $1.48 \pm 0.19$      & {[WMO2009] For-1581} & $-0.57 \pm 0.08$ & $1.83 \pm 0.29$\\
{[WMO2009] For-2280} & $-0.79 \pm 0.05$ & $1.77 \pm 0.17$         & {[LHT2010] BL233} & $-0.56 \pm 0.12$ & $1.83 \pm 0.17$\\
{[LDH2014] Fnx-mem0556} & $-0.79 \pm 0.09$ & $2.30 \pm 0.17$      & {[WMO2009] For-2026} & $-0.54 \pm 0.06$ & $1.90 \pm 0.13$\\
{[LHT2010] BL084} & $-0.78 \pm 0.10$ & $1.70 \pm 0.19$            & {2MASS J02393808-3437062} & $-0.54 \pm 0.08$ & $1.90 \pm 0.15$\\
{[WMO2006] F01-23} & $-0.78 \pm 0.06$ & $1.62 \pm 0.11$           & {[WMO2009] For-1120} & $-0.54 \pm 0.08$ & $2.00 \pm 0.13$\\
{[LDH2014] Fnx-mem0607} & $-0.77 \pm 0.09$ & $1.66 \pm 0.16$      & {2MASS J02394195-3430361} & $-0.54 \pm 0.06$ & $2.00 \pm 0.14$\\
{[LDH2014] Fnx-mem0572} & $-0.77 \pm 0.04$ & $1.78 \pm 0.21$      & {2MASS J02384018-3439121} & $-0.53 \pm 0.07$ & $1.96 \pm 0.19$\\
{2MASS J02390434-3425190} & $-0.76 \pm 0.04$ & $1.66 \pm 0.12$    & {2MASS J02390819-3436537} & $-0.52 \pm 0.08$ & $1.93 \pm 0.15$\\
{[LDH2014] Fnx-mem0532} & $-0.76 \pm 0.11$ & $1.52 \pm 0.25$      & {2MASS J02392769-3437487} & $-0.51 \pm 0.08$ & $2.01 \pm 0.12$\\
{[LDH2014] Fnx-mem0543} & $-0.75 \pm 0.10$ & $1.66 \pm 0.13$      & {2MASS J02395604-3424106} & $-0.51 \pm 0.06$ & $2.01 \pm 0.13$\\
{[LHT2010] BL311} & $-0.75 \pm 0.04$ & $1.78 \pm 0.19$            & {2MASS J02392483-3434383} & $-0.50 \pm 0.05$ & $2.00 \pm 0.19$\\
{[LDH2014] Fnx-mem0539} & $-0.74 \pm 0.16$ & $2.01 \pm 0.21$      & {WEL  60} & $-0.49 \pm 0.09$ & $1.98 \pm 0.14$\\
{[LHT2010] BL138} & $-0.74 \pm 0.03$ & $1.81 \pm 0.11$            & {2MASS J02394309-3440186} & $-0.49 \pm 0.08$ & $1.92 \pm 0.13$\\
{FBW J024006.1-342852} & $-0.74 \pm 0.09$ & $1.70 \pm 0.13$       & {2MASS J02393412-3433096} & $-0.48 \pm 0.08$ & $2.04 \pm 0.12$\\
{[WMO2006] F01-6} & $-0.73 \pm 0.07$ & $1.91 \pm 0.24$            & {[LDH2014] Fnx-mem0522} & $-0.47 \pm 0.06$ & $1.96 \pm 0.17$\\
{[LHT2010] BL104} & $-0.73 \pm 0.09$ & $1.81 \pm 0.19$            & {[LHT2010] BL298} & $-0.46 \pm 0.06$ & $1.93 \pm 0.18$\\
{[WMO2009] For-0952} & $-0.73 \pm 0.04$ & $1.74 \pm 0.20$         & {2MASS J02395732-3431211} & $-0.43 \pm 0.07$ & $2.10 \pm 0.22$\\
{2MASS J02382132-3436180} & $-0.73 \pm 0.12$ & $1.64 \pm 0.19$    & {[WMO2009] For-0949} & $-0.42 \pm 0.08$ & $1.99 \pm 0.15$\\
{[LHT2010] BL146} & $-0.73 \pm 0.08$ & $1.77 \pm 0.17$            & {[WMO2006] F01-16} & $-0.38 \pm 0.06$ & $2.02 \pm 0.18$\\
{[WMO2009] For-1208} & $-0.72 \pm 0.06$ & $1.90 \pm 0.17$         & {[LHT2010] BL163} & $-0.35 \pm 0.07$ & $1.86 \pm 0.15$\\
{[WMO2009] For-1117} & $-0.72 \pm 0.06$ & $1.90 \pm 0.19$         & {2MASS J02393084-3435451} & $-0.35 \pm 0.06$ & $2.20 \pm 0.20$\\
{GB07 Fnx24} & $-0.72 \pm 0.08$ & $1.51 \pm 0.13$                 & {[MOW91]  21} & $-0.31 \pm 0.11$ & $2.00 \pm 0.25$\\
{[LDH2014] Fnx-mem0754} & $-0.72 \pm 0.05$ & $1.80 \pm 0.17$      \\
 \hline                                   
\end{tabular}
\end{table*}   

\bibliography{sample63}{}
\bibliographystyle{aasjournal}



\end{document}